\documentclass[twocolumn]{aastex631}

\usepackage{acronym}
\usepackage{amsmath}
\usepackage{bm}

\shorttitle{GRB 170817A In Full View}
\shortauthors{Ryan et al.}

\newcommand{\afterglowpy}{{\texttt{afterglowpy}}}
\newcommand{\emcee}{{\texttt{emcee}}}
\newcommand{\gwgrb}{GW170817}

\newcommand{\thobs}{\ensuremath{\theta_{\mathrm{obs}}}}
\newcommand{\tobs}{\ensuremath{t_{\mathrm{obs}}}}
\newcommand{\nuobs}{\ensuremath{\nu_{\mathrm{obs}}}}

\newcommand{\xobs}{\ensuremath{x_{\mathrm{obs}}}}
\newcommand{\yobs}{\ensuremath{y_{\mathrm{obs}}}}
\newcommand{\zobs}{\ensuremath{z_{\mathrm{obs}}}}
\newcommand{\xobsang}{\ensuremath{\tilde{x}_{\mathrm{obs}}}}
\newcommand{\yobsang}{\ensuremath{\tilde{y}_{\mathrm{obs}}}}

\newcommand{\xcen}{\ensuremath{\tilde{x}_{\mathrm{c}}}}
\newcommand{\ycen}{\ensuremath{\tilde{y}_{\mathrm{c}}}}

\newcommand{\thC}{\ensuremath{\theta_{C}}}
\newcommand{\thW}{\ensuremath{\theta_{W}}}
\newcommand{\epse}{\ensuremath{\epsilon_{e}}}
\newcommand{\epsB}{\ensuremath{\epsilon_{B}}}
\newcommand{\xiN}{\ensuremath{\xi_{N}}}
\newcommand{\LX}{\ensuremath{L_{X, 38}}}
\newcommand{\refdec}{\ensuremath{\mathrm{DEC}_0}}
\newcommand{\refra}{\ensuremath{\mathrm{RA}_0}}
\newcommand{\posang}{\ensuremath{\mathrm{PA}}}

\newcommand{\Etot}{\ensuremath{E_{\mathrm{tot}}}}
\newcommand{\RA}{\ensuremath{\mathrm{RA}}}
\newcommand{\dec}{\ensuremath{\mathrm{Dec}}}
\newcommand{\PA}{\ensuremath{\mathrm{PA}}}

\newcommand{\MM}{\mathcal{M}}

\newcommand{\Planck}{{Planck}}
\newcommand{\Chandra}{{Chandra}}
\newcommand{\HST}{{HST}}
\newcommand{\JWST}{{JWST}}

\newcommand{\citetNewScaling}{{van Eerten and Ryan (2023, in prep)}}

\newcommand{\CitetNewScaling}{{Van Eerten and Ryan (2023, in prep)}}

\acrodef{GRB}{gamma-ray burst}
\acrodef{VLBI}{very-long-baseline interferometry}
\acrodef{GW}{gravitational wave}
\acrodef{LSC}{LIGO Scientific Collaboration}

\graphicspath{{figures/}}

\begin{document}

\title{Modelling of Long-Term 
Afterglow Counterparts to Gravitational Wave Events:\\ The Full View of GRB 170817A} 

\author[0000-0001-9068-7157]{Geoffrey Ryan}
\affiliation{Perimeter Institute for Theoretical Physics, Waterloo, Ontario N2L 2Y5, Canada}

\author[0000-0002-8680-8718]{Hendrik van Eerten}
\affiliation{Department of Physics, University of Bath, Claverton Down, Bath, BA2 7AY, UK}

\author[0000-0002-1869-7817]{Eleonora Troja}
\affiliation{Department of Physics, University of Rome ``Tor Vergata'', via della Ricerca Scientifica 1, I-00133 Rome, Italy}

\author[0000-0003-4159-3984]{Luigi Piro}
\affiliation{INAF, Istituto di Astrofisica e Planetologia Spaziali, via del Fosso del Cavaliere 100, I-00133 Rome, Italy}

\author[0000-0002-9700-0036]{Brendan O'Connor}
\altaffiliation{McWilliams Fellow}
\affiliation{McWilliams Center for Cosmology, Department of Physics, Carnegie Mellon University, Pittsburgh, PA 15213, USA}

\author[0000-0003-4631-1528]{Roberto Ricci}
\affiliation{Istituto Nazione di Ricerca Metrologica, Strada delle Cacce 91, I-10135 Turin, Italy}
\affiliation{INAF-Istituto di Radioastronomia, Via Gobetti 101, I-40129 Bologna, Italy}

\correspondingauthor{Geoffrey Ryan}
\email{gryan@perimeterinstitute.ca}



\begin{abstract}
The arrival of gravitational wave astronomy and a growing number of time-domain focused observatories are set to lead to a increasing number of detections of short gamma-ray bursts (GRBs) launched with a moderate inclination to Earth.
Being nearby events, these are also prime candidates for very long-term follow-up campaigns and very-long-baseline interferometry (VLBI), which has implications for multi-messenger modelling, data analysis, and statistical inference methods applied to these sources. 
Here we present a comprehensive modelling update that directly incorporates into \afterglowpy{} astrometric observations of the GRB position, Poissonian statistics for faint sources, and modelling of a trans-relativistic population of electrons. 
We use the revolutionary event \gwgrb{} to demonstrate the impact of these extensions both for the best-fit physics parameters and model selection methods that assess the
statistical significance of additional late-time emission components.
By including in our analysis the latest \Chandra{} X-ray  observations of GRB 170817A, we find only weak evidence ($\lesssim$2 $\sigma$) for a new emission component at late times, which makes for a slightly more natural fit to the centroid evolution and prediction for the external medium density.

\end{abstract}



\section{Introduction} \label{sec:intro}

\Ac{GRB} 170817A has revolutionized our understanding of short GRBs (i.e. those GRBs with prompt duration less than about 2 seconds) and their afterglows. Famous as the first-ever electromagnetic counterpart to a direct detection of gravitational waves (GWs) from a neutron star merger \citep{Abbott:2017multi-messenger} and occurring alongside an exquisitely well-sampled kilonova, GRB 170817A was both a nearby event (at $40$ Mpc distance) and observed at a significant angle relative to the axis of its jetted prompt and afterglow emission. This orientation relative to the merger plane was already apparent from the GW detection itself \citep{Abbott:2017discovery}, but once it was established that the GRB and afterglow were indeed the products of a collimated relativistic jet of plasma 
(see e.g. \citealt{Troja:2017, Lazzati:2018, Troja:2018GRB170817A, Margutti:2018, Mooley:2018, Troja:2019ab, Hajela:2019aa}), the orientation angle $\thobs$ became a key ingredient in the event's modelling and analysis. A consistent outcome of afterglow modelling of GRB 170817A is that, orientation aside, the afterglow model parameters are in line with those obtained for the general sample of short GRBs \citep{Troja:2019ab,Wu:2019}. This lends credence to the notion that GW170817 / GRB 170817A establishes that, in general, short GRBs are the products of neutron star mergers.

However, the atypical orientation of the jet did reveal an important morphological feature of GRB afterglow jets, which is that they appear to possess a rich lateral structure in their distribution of outflow energies and Lorentz factors (this helps explain the prompt emission as well, see e.g. \citealt{Troja:2017, IokaNakamura:2019, BeniaminiNakar:2019, Matsumoto:2019}). GRB 170817A and other GRBs that have subsequently been argued to have been observed at a significant angle (e.g. GRB 150101B, \citealt{Troja:2018}) are modelled best by an outflow profile where energy drops with increasing angle from the jet axis, in practice this profile is often taken to be a \emph{Gaussian} or a \emph{power law}. This allows for light curves (seen from radio to X-rays for GRB 170817A) that rise steadily until a late turnover point, while maintaining the collimated nature of the outflow demanded \citep{Troja:2018GRB170817A} by the steep post turnover decay slope of the light curve and the centroid motions revealed by \ac{VLBI} radio observations of the jet at late times (for VLBI analysis see \citealt{Mooley:2018, Hotokezaka:2019aa, Ghirlanda:2019, Fernandez:2021to, Mooley:2022vs, Govreen-Segal:2023aa}, with \citealt{Zrake:2018} providing a counter point about the implications for collimated flow and \citealt{Nedora:2023ab} considering the impact of a kilonova afterglow). GW detections do not select for on-axis GRBs and are limited to the distance horizon of the GW detectors. This means that future EM counterparts discovered during LIGO-Virgo-KAGRA observing run 4 (O4) and subsequent runs will be nearby and highly probable to be observed at significant angle. If GRB 170817A is representative of future afterglow EM counterparts, structured jet models will therefore continue to play a key role in their understanding.

The fact that GRB 170817A is still actively being monitored six years after the initial GW event raises an important point that motivates the remainder of this paper: nearby off-axis structured jet GRBs might remain visible for a long period of time, during which they will transit into non-relativistic flow. This invites further refining of the structured jet afterglow model and of our methods to extract the maximum amount of information from multi-messenger observations in the context of long-term broadband monitoring campaigns. In particular, we address the following points:
\begin{enumerate}
\item It is well known that standard GRB afterglow models assume ultra-relativistic plasma for the purpose of computing the afterglow synchrotron emission, even when the jet's bulk motion becomes non-relativistic. Structured jet models should properly incorporate the transition into non-relativistic into both the jet dynamics and the synchrotron emission prescription.
\item The multi-messenger nature of GW-EM observations offers a great opportunity to drive the statistical inference jointly with completely different channels of information. High precision astrometric mesaurements, such as from VLBI, Hubble Space Telescope (\HST{}), and \JWST{}, offer another channel uniquely sensitive to the system geometry in addition to GW detections and broadband light curve photometry.
\item Even when nearby, the emission of EM counterpart GRBs is likely to be very faint during late stages. Modelling should therefore properly account for the nature of the background and low photon counts, and Poissonian statistics needs to be applied.
\item With its persistent X-ray emission, GRB 170817A already offers a tantalizing glimpse of potential features of the system only showing up at late times. To quantitatively assess the significance of deviations from the basic model at late times (or any time, for that matter), the appropriate statistical inference techniques should be utilized for the purpose of model selection.
\end{enumerate}

This paper is organized as follows. In Section \ref{sec:physics} we review the structured jet forward shock model and discuss its extensions to include astrometric observables and the late-time electron population.  In Section \ref{sec:data} we discuss our updated data analysis pipeline.  Section \ref{sec:170817} applies our methods to \gwgrb{}, and Section \ref{sec:discussion} discusses application to future events.

\section{Jet Afterglow Model} \label{sec:physics}

The principal component of a \ac{GRB}'s non-thermal afterglow is synchrotron radiation emitted from a relativistic blast wave launched by the \ac{GRB} progenitor \citep[e.g.][]{GranotSari:2002}. 

We use the \afterglowpy{}\footnote{Available at \url{https://pypi.org/project/afterglowpy/}} software package to compute the forward shock emission of the GRB afterglow \citep{Ryan:2020aa}.  We consider ``structured'' jets with non-trivial angular energy profiles $E(\theta)$, typically one of:
\begin{align}
    E(\theta) &= E_0 \exp\left(-\frac{1}{2}\left(\frac{\theta}{\thC}\right)^2\right) & &\text{Gaussian,}\label{eq:GaussianJet} \\*
    E(\theta) &= E_0 \left(1 + \frac{1}{b}\left(\frac{\theta}{\thC}\right)^2\right)^{-b/2} & &\text{Power-Law.}\label{eq:PowerlawJet}
\end{align}
Here $E_0$ is the on-axis isotropic equivalent energy of the blast wave, $\thC$ is the opening angle of the jet core, $b$ is the power-law index of the jet ``wings'' outside the core, and $\theta$ is the angle measured from the jet axis. Both these profiles are also truncated at an outer angle $\thW$ which is a free parameter.

Standard afterglow modelling (e.g. \citealt{GranotSari:2002}) often relies on the assumption that the shock-acceleration process creates a population of non-thermal electrons residing fully at relativistic energies, leading to a power law distribution in energy with a slope $-p$ (canonically, $p$ is assumed to lie around $2.2 - 2.6$, e.g. \citealt{Curran:2010}).  Additionally, it is commonly assumed that this applies to \emph{all} available electrons, setting participation fraction $\xiN \equiv 1$. These electrons are set to contain a fraction $\epse$ of the post-shock internal energy density of the plasma (typically, $\epse \sim 0.1$ \cite{Beniamini:2017wu, Duncan:2023aa}). The accelerated particles produce synchrotron emission when interacting with a magnetic field containing a fraction $\epsB$ of the available post-shock energy density.

In the following sections we review the effect of viewing and jet geometry on the GRB afterglow light curve (Section \ref{sec:geom}), discuss the \emph{Deep Newtonian limit} addition to the basic forward shock model that better captures the late-time behaviour of GRB afterglows (Section \ref{sec:deepnewtonian}), and discuss proper motion of the afterglow radio centroid (Section \ref{sec:centroidmodel}).  The \afterglowpy{} code \texttt{v0.8.0} includes these new features and has also undergone some numerical improvements to improve speed and robustness. In particular, the original Romberg integrator has been replaced with a custom adaptive quadrature routine based on the CADRE scheme \citep{deBoor:1971aa, Gonnet:2010aa}.  This integrator gives robust error estimation and minimizes the number of evaluations required to produce an afterglow flux within the required tolerance, typically set to be $\lesssim 1 \%$.

\subsection{How Geometry Changes The Light Curve} \label{sec:geom}

In the standard \ac{GRB} afterglow scenario of a top-hat jet viewed on-axis ($\theta_{obs} = 0$), relativistic beaming makes it impossible to distinguish a jet from a spherical flow until the onset of the \emph{jet break}: the time when the entire jet surface is in view, the jet begins to spread laterally, and the observed flux drops quickly. The time of this jet break is determined by the jet opening angle $\thC$, energy $E_0$, and circumburst density $n$: $t_{jb}(\thobs=0) \sim (E_0/n)^{1/3} \thC^{8/3}$ \citep{Rhoads:1999, Sari:1999, vanEerten:2010}.  

After the jet break the jet spreads, decelerates, and approaches the non-relativistic spherical Sedov regime \citep{Zhang:2009ApJ}. The emitted flux in the Sedov regime depends on the total energy in the jet $\Etot \sim E_0 \thC^2 \ll E_0$.  Wide jets (with large $\thC$, and the same $E_0$) are brighter at late times as they have larger total energy.  Late-time calorimetry, often performed in the radio, seeks to measure $\Etot$ and $\thC$ to constrain the overall energetics of the GRB central engine \citep[e.g.][]{Frail:2000aa}.  Geometry (here, just the width of the jet) affects these afterglows in two ways: wider jets have later jet breaks ($t_{jb} \propto \thC^{8/3}$) and brighter emission in the Sedov phase than narrower jets (at fixed $E_0$).

\begin{figure}
    \includegraphics[width=\columnwidth]{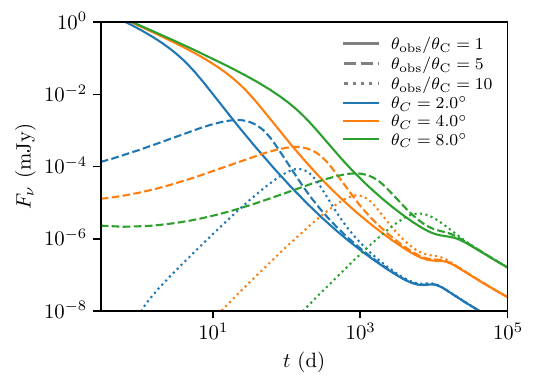}
    \caption{\label{fig:lcAngles} Fiducial optical GRB afterglow light curves of a Gaussian structured jet for a variety of opening angles $\thC$ and viewing angle to opening angle ratios $\thobs / \thC$ at $\nuobs = 10^{14}$ Hz.  Colors denote opening angle (blue: $\thC = 2^\circ$, orange: $4^\circ$, green: $8^\circ$), line styles the viewing angle ratio (solid: $\thobs/\thC = 1$, dashed: 5, dotted: 10). Parameters for the jet unless otherwise stated are chosen to emulate \gwgrb{}: $\thobs = 19.4^\circ$, $E_0 = 4.8\times 10^{53}$ erg, $\thC = 3.2^\circ$, $\thW = 4.9 \thC$, $n_0 = 2.4 \times 10^{-3} \mathrm{cm}^{-3}$, $p = 2.13$, $\epse = 1.9\times 10^{-3}$, $\epsB = 5.75\times 10^{-4}$, $\xiN = 1.0$, $d_L = 40$ Mpc, $z = 0.0973$. Jets with the same ratio $\thobs / \thC$ share the same initial power-law slope.  At late times (after the jet break) light curves from jets with the same opening angle converge, erasing the effect of the viewing angle.  The counter-jet (the jet directed away from the Earth) appears at very late times, $\sim 10^4$ days. }
\end{figure}

The picture is somewhat complicated when jets are allowed to have angular structure and be viewed off the jet axis \citep{Meszaros:1998aa, Rossi:2002, KumarGranot:2003}. Relativistic beaming may cause the early light curve to display a shallow decay or even a rise (as in the case of \gwgrb{}), with a particular slope determined by the ratio $\thobs / \thC$ and the specific angular structure $E(\theta)$ (\citet{Ryan:2020aa}, for subsequent discussion see also \citet{Beniamini:2020ab, Nakar:2021up}). 
The jet break is delayed relative to the on-axis case by a factor that also depends on $\thobs / \thC$ (Equation 4 of \citealt{vanEerten:2010}; see also \citealt{Ryan:2015aa}).  The structure profile can also affect the sharpness of the jet break \citep{Lamb:2021aa} or in some cases even steepen the early light curve \citep{Beniamini:2022ty, OConnor:2023aa}. 

Following the jet break the jet spreads, decelerates to sub-relativistic velocities, and the viewing angle dependence of the emission disappears. As in the on-axis case the jet approaches the spherical Sedov phase with a flux determined by the total energy $\Etot = 2\pi \int_0^{\thW} E(\theta) \sin(\theta) d\theta$ instead of just $E_0$. Geometry (now both the viewing angle $\thobs$ and jet width $\thC$) affects these light curves in three ways: the initial (relativistic, pre-jet-break) slope is determined by $\thobs / \thC$ and $E(\theta)$, the jet break occurs at a time $t_{jb} \sim (\thobs + \thC)^{8/3}$, and the Sedov phase emission increases with $\Etot$ (and hence $\thC$).

Figure \ref{fig:lcAngles} shows the afterglow light curves from a Gaussian structured jet of three different opening angles $\thC$ and at three different viewing angles to illustrate these behaviours. 
The early light curves are drastically different, primarily controlled by the ratio $\thobs / \thC$, which sets the light curve slope as well as the delay of the jet break. At late times the light curves at different viewing angles all collapse onto one curve, determined by the jet opening angle.  Wider jets correspond to brighter emission at late times.

When an off-axis GRB afterglow is observed, the pre-jet-break light curve constrains $\thobs / \thC$ and $E(\theta)$, while the post-jet-break Sedov light curve constrains $\Etot$.  Both phases must be observed, and $E(\theta)$ assumed, to be able to determine both $\thobs$ and $\thC$ independently from the light curve alone. If the afterglow emission is contaminated at late times by an additional source, or if the afterglow model used is inaccurate in the Sedov phase, then neither $\thobs$ nor $\thC$ will be correctly measured.

\subsection{Modelling of synchrotron emission and the Deep Newtonian Regime}
\label{sec:deepnewtonian}

\begin{figure}
    \includegraphics[width=\columnwidth]{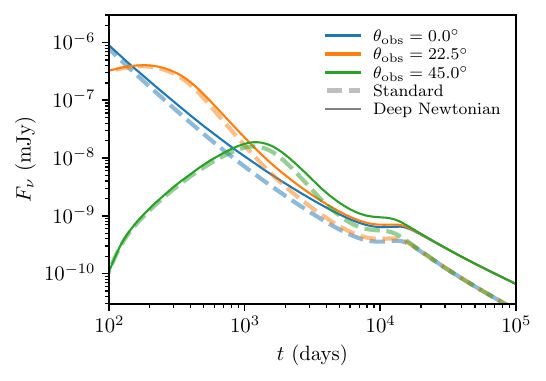}
    \caption{\label{fig:dnXrayLC} Afterglow X-ray light curves with (solid lines) and without (pale dashed line) the Deep Newtonian regime at different viewing angles (colors; blue: $\thobs = 0^\circ$, orange: $22.5^\circ$, green: $45^\circ$) at $\nuobs=5$ keV. Unless otherwise stated the jet parameters are identical to those in Figure \ref{fig:lcAngles}. With these parameters, $\epse = 1.9\times10^{-3}$ and $\xiN = 1.0$, the Deep Newtonian effects begin to alter the light curve when the jet is still trans-relativistic, before the transition to the Sedov phase is complete and the counter-jet appears.}
\end{figure}

Our model has been expanded over the standard approach to afterglow emission  modelling in order to deal with emission from blast waves evolving over a long period of time: specifically, including the \emph{Deep Newtonian} regime of \citet{SironiGiannios:2013} and \citet{HuangCheng:2003}). We briefly review the rationale for this approach and discuss its implementation and implications below. Including the Deep Newtonian regime will lead to a relative increase in flux at late times by a factor of a few and a shallower late-time light curve slope (as shown in Figure \ref{fig:dnXrayLC}).

The standard assumption, utilized by \afterglowpy{}, is that at the forward shock a population of electrons are accelerated into a non-thermal power law distribution $n'_e \sim \gamma_e^{-p}$, with $\gamma_e$ the electron Lorentz factor and the index $p > 2$.  This distribution is bounded from below by the injection Lorentz factor $\gamma_m$, given by \citep[e.g.][]{Sari:1998}:
\begin{equation}
    \gamma_m = \frac{p-2}{p-1} \frac{\epse e' }{\xiN n' m_e c^2}\ .
\label{eq:gamma_m}
\end{equation}
Here $n'$ is the proton number density in the shocked fluid rest frame, $e'$ is the thermal energy density in the same frame, $\xiN$ is the fraction of electrons that get accelerated, $\epse$ is the fraction of energy in the non-thermal electrons, and $m_e$ and $c$ are the electron mass and the speed of light, respectively.  Equation \ref{eq:gamma_m} results simply from integrating $\gamma_e m_e c^2\times n'_e(\gamma_e)$ from $\gamma_m$ to $\infty$ and requiring the result equal $\epse e'$. Importantly, this procedure assumes throughout that $\gamma_e \gg 1$.  

\citet{SironiGiannios:2013} point out that when the forward shock becomes non-relativistic (but is still a strong shock) Equation \ref{eq:gamma_m} breaks down and can produce $\gamma_m < 1$.  At this point, the true non-thermal electron distribution is a power law in momentum, not energy, and while the lowest energy electrons will be non-relativistic, the population maintains a relativistic tail which can continue to produce synchrotron radiation.

This behaviour is labeled the \emph{Deep Newtonian} regime \citep{SironiGiannios:2013, HuangCheng:2003}, and modifies both the spectral and temporal evolution of GRB afterglows at late times, approaching the Sedov stage.  Since GW-GRB counterparts like \gwgrb{} will tend to be nearby and subject to intense long-term follow-up we have the opportunity to observe their trans- and non-relativistic evolution in detail. Including these effects is important to their accurate modelling.

We have implemented the Deep Newtonian regime in \texttt{v0.8.0} of \afterglowpy{}, following the recipe of \citet{SironiGiannios:2013}.  Since the synchrotron-emitting electrons are still relativistic, much of the standard treatment can remain intact.  We first compute $\gamma_m$ as normal following Equation \ref{eq:gamma_m}, then bound its value from below by 1, and finally take the number density of emitters (proportional to the emission and absorption coefficients) to be $(p-2)\epsilon_e e' / (\gamma_m m_e c^2)$.  This appropriately returns the number density of relativistic electrons, equal to $(p-1) \xiN n'$, only if $\gamma_m > 1$.  See \citetNewScaling{} for a more detailed discussion of the flux scaling relations in this regime.

Figure \ref{fig:dnXrayLC} shows the effect of the Deep Newtonian regime on fiducial GRB afterglow light curves.  For these parameters the Deep Newtonian limit becomes relevant when the jet has Lorentz factor $\gamma \approx 3.5$. For on-axis bursts the mild flux enhancement begins well after the jet break, during the slow approach to the Sedov phase. However, if the viewing angle is increased the tip of the jet will come into view later and with a smaller Lorentz factor, until the point where the Deep Newtonian enhancement appears even during the jet break at $\thobs = 45^\circ$. Once the Deep Newtonian regime becomes apparent, the difference between the ``standard'' and corrected light curves grows with time, leading to a factor $\sim$few increase to the flux $\sim 30$ years after the burst in the fiducial afterglow model. The departure between the models would be delayed and lessened for bursts with larger $\epse$ or smaller $\xiN$ than our fiducial values.

\subsection{Centroid Motion} \label{sec:centroidmodel}

\begin{figure*}
    \includegraphics[width=\textwidth]{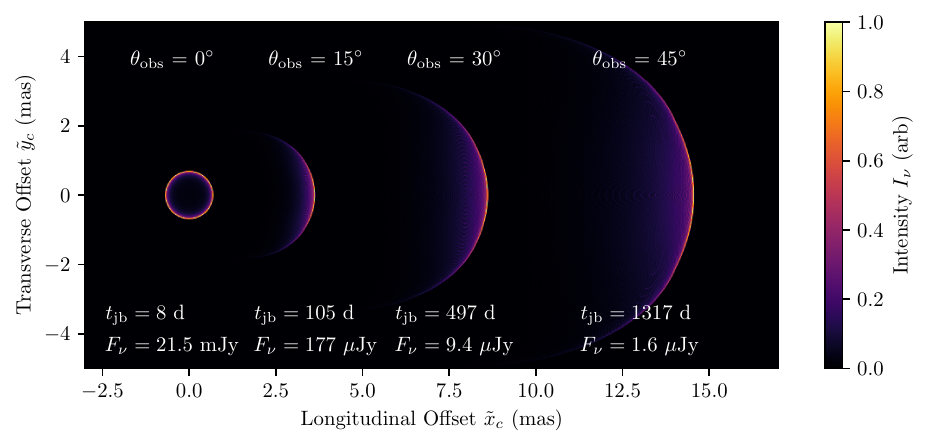}
    \caption{\label{fig:vlbiImages} GRB afterglow radio images for a \gwgrb{}-like jet at different viewing angles $\thobs$ at the time of the jet break. Shown is the specific intensity $I_\nu$ at $\nu = 5$ GHz of the primary jet, scaled so each image has the same maximum intensity.  Doppler beaming produces a limb-brightened image, with emission dominated by the high Lorentz factor material near the jet core. Unless otherwise stated the jet parameters are identical to those in Figure \ref{fig:lcAngles}.}
\end{figure*}

A \ac{GRB} blast wave can evolve over light-year length scales and remain visible.  For \acp{GRB} $\lesssim$ 100 Mpc distant the afterglow image on the sky will have a $\sim 1$ mas angular scale, a scale probeable with radio \ac{VLBI} and high precision optical astrometry (e.g. \HST{} and \JWST{}). This was successfully done for \gwgrb{} \citep{Mooley:2018, Ghirlanda:2019, Mooley:2022vs}, yielding critical geometric information. We have therefore expanded \afterglowpy{} to allow for modelling of the centroid evolution.  

The afterglow image centroid position and effective size may be computed by taking moments of the specific intensity distribution in the observer's sky. This same technique has been used by \citet{Zrake:2018, Fernandez:2021to, Mooley:2022vs, Nedora:2023aa}. We consider a relativistic jet (and optionally its counter-jet) and a Cartesian coordinate system centered on the jet's origin with $\mathbf{\hat{z}}$-axis aligned along the jet axis.  A distant observer is located at an angle $\thobs$ from $\mathbf{\hat{z}}$ towards $\mathbf{\hat{x}}$.  The observed flux in the optically thin limit may be computed as
\begin{equation}
    F_\nu(\tobs, \nuobs) = \frac{1+z}{4\pi d_L^2}\int d\Omega d R \ R^2 P_\nu\ ,
\end{equation}
where $P_\nu$ is the isotropic-equivalent power density emitted towards the observer, $R$ is the radial coordinate from the origin, $d\Omega$ is the solid angle element, $z$ is the redshift, and $d_L$ the luminosity distance.

We orient the distant observer's local coordinate system ($\xobs, \yobs, \zobs$) such that their $\hat{\mathbf{z}}_{\mathrm{obs}}$-axis is oriented away from the jet origin, \xobs{}-\yobs{} forms an image plane, and the jet head moves in the $+\xobs$ direction.  Blast wave elements at location $(x, y, z)$ will then occur at proper distance $\xobs$, $\yobs$ relative to the observer:
\begin{align}
     \xobs &= -x \cos \thobs + z \sin \thobs  \ ,\\*
     \yobs &= -y\ .
\end{align}
In the observer's sky, the same blast wave element will appear at an angle $\xobsang = \xobs / d_A$ along $\xobs$ from the jet origin, where $d_A = d_L / (1+z)^2$ is the angular diameter distance. Similarly $\yobsang = \yobs / d_A$.

Figure \ref{fig:vlbiImages} shows the images of a \gwgrb{}-like jet at $\thobs = $ $0^\circ$, $15^\circ$, $30^\circ$, and $45^\circ$, each at the time of the jet break.  

Defining an intensity-averaged quantity $\langle f \rangle$ as:
\begin{equation}
    \left \langle f \right \rangle = \left .  \int d\Omega dR\ f\ R^2 P_\nu\quad \middle / \quad  \int d\Omega dR\ R^2 P_\nu \right .
\end{equation}
The centroid of emission at time $\tobs$ and frequency $\nuobs$ is then located at:
\begin{equation}
    \xcen = \left\langle \xobsang \right\rangle, \qquad \ycen = \left\langle \yobsang \right\rangle
\end{equation}
Higher order moments are similarly constructed.  The Gaussian-equivalent image size and orientation can be constructed from:
\begin{align}
    \tilde{\sigma}^2_x &= \big \langle \left(\xobsang - \xcen\right)^2 \big\rangle  \\*
    \tilde{\sigma}^2_{xy} &= \big \langle \left(\xobsang-\xcen\right)\left( \yobsang-\ycen\right) \big\rangle \\*
    \tilde{\sigma}^2_y &= \big\langle \left(\yobsang-\ycen\right)^2 \big \rangle
\end{align}
Calculation of the intensity-weighted image moments has been added to the \afterglowpy{} package \texttt{v0.8.0} \citep{Ryan:2020aa}.  The \afterglowpy{} jet models are axisymmetric by construction, so $\ycen = \tilde{\sigma}^2_{xy} = 0$.

\begin{figure*}
    \begin{center}
    \begin{tabular}{cc}
    \includegraphics[width=0.48\textwidth]{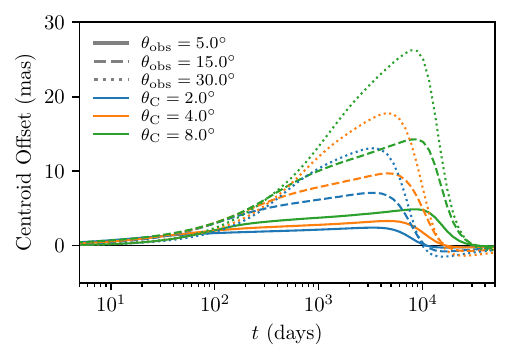} & \includegraphics[width=0.48\textwidth]{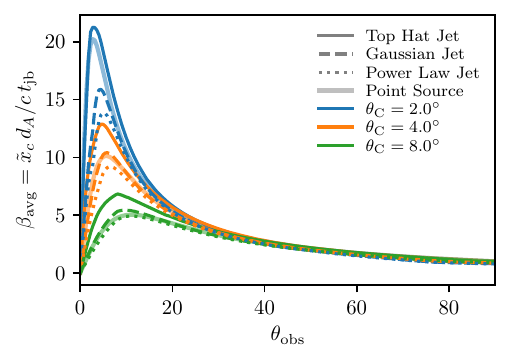}    
    \end{tabular}
    \end{center}
    \caption{\label{fig:offset} Left panel: Fiducial GRB afterglow centroid offsets of a Gaussian structured jet for a variety of opening angles $\thC$ and viewing angles $\thobs$.  Line styles denote opening angle, colors the viewing angle.
    Right Panel: The apparent velocity of the GRB afterglow centroid measured from the initial burst to the jet break as a function of viewing angle. Line styles denote jet structure profile, line colors denote jet opening angle. Unless otherwise stated the jet parameters are identical to those in Figure \ref{fig:lcAngles}.}
\end{figure*}

Figure \ref{fig:offset} (left panel) shows the observed angular offset of the afterglow centroid as a function of time for a Gaussian \gwgrb{}-like jet.  The offset increases quickly with time in the initial relativistic evolution, slows as the jet becomes transrelativistic, before falling back to the origin at very late times as the counter-jet comes into view.  During the relativistic motion of the jet, before the jet break, the apparent velocity can be superluminal.  

The magnitude of the offset is a strong function of the viewing angle.  The offset is maximal for some viewing angle, typically $\lesssim 20^\circ$, depending on the time of observation.  Wider jets display larger offsets.  As with the light curve, the  evolution of the centroid offset slows as both the viewing angle and opening angle increase. We note the long-term non-relativistic behaviour appears to depend on the model of jet spreading used, see \citet{Fernandez:2021to} for comparison.

The observed magnitude of the centroid offset $\xcen$ depends inversely on the angular-diameter distance to the source $d_A$ as well as on the energy of the GRB jet $E_0$ and the density of the circumburst environment $n_0$. Because of this dependence on $E_0$ and $n_0$, it is difficult to extract clean geometrical information from a measurement of the centroid offset alone.  Assuming the distance is known, a large offset could imply a particular viewing angle \emph{or} an underdense medium, for instance.  

A better diagnostic is the average apparent velocity of the centroid from launch to the time of the jet break $t_{\mathrm{jb}}$:
\begin{equation}
    \beta_{\mathrm{avg}} \equiv x_c(t_{\mathrm{jb}}) / c t_{\mathrm{jb}}\ .
\end{equation}  
The jet break is a time measure intrinsic to the GRB jet and inherits the same dependence on $E_0$ and $n_0$ as the proper offset $x_c$.  The average velocity $\beta_{\mathrm{avg}}$ then depends \emph{only} on the viewing angle and jet structure, and holds much of the constraining power of the centroid motion. \CitetNewScaling{} contains a detailed discussion of the scaling relations for the afterglow flux, the centroid offset, and the scale invariance of $\beta_{\mathrm{avg}}$.  Basing the centroid measure on the jet break time is also useful observationally: for off-axis jets $\thobs \gg \thC$ the light curve rises with time and peaks at $t = t_{\mathrm{jb}}$, making this the best (or last, if the jet is closer to on-axis) time to attempt a VLBI astrometric measurement. 

The right panel of Figure \ref{fig:offset} shows $\beta_{\mathrm{avg}}$ as a function of the viewing angle for a selection of jet structure models and widths. The jet break time was computed using the fitting formula \citet{Ryan:2020aa} (Equation 39).  We have also computed $\beta_{\mathrm{avg}}$ for a simple point source moving with constant Lorentz factor $\Gamma = 1 / \thC$, the characteristic velocity at the jet break.  For any particular jet the apparent velocity peaks at approximately $\thobs \sim \thC$.  This agrees with the simple point source model: superluminal apparent velocity is maximal for viewing at $\thobs \sim 1/\Gamma$, and the jet break occurs when $\Gamma \sim 1 / \thC$.  In general, $\beta_{\mathrm{avg}}$ is highly sensitive to the jet model (width and structure profile) when viewed at $\thobs \lesssim 25^\circ$  \citep[see also][]{Fernandez:2021to}. On the other hand,  at large angles ($\thobs \gtrsim 25^\circ$) all models converge on the point source approximation $\beta_{\mathrm{avg}} \approx \cot( \thobs / 2)$.  This makes $\beta_{\mathrm{avg}}$ in principle a very clean proxy for $\thobs$, although the weak dependence of $\beta_{\mathrm{avg}}$ on $\thobs$ when $\thobs \gg 20^\circ$ may make this challenging in practice.

Measurements of $\beta_{\mathrm{avg}}$ are conceptually simple, even if challenging observationally, and yield immediate geometric information independent of the jet energy, density, etc.  For \gwgrb{} the closest measure of $\beta_{\mathrm{avg}}$ comes from the $8$ and $206$ day observations yielding a value of $4.7\pm 0.6$ \citep{Ghirlanda:2019, Mooley:2022vs}.  Examining the $\beta_{\mathrm{avg}}$ panel of Figure \ref{fig:offset}, most models cross $\beta_{\mathrm{avg}} = 4.7$ at angles $\lesssim 10^\circ$ or $\sim 20^\circ$.  It seems jets wider than $\sim 8^\circ$ do not reach $\beta_{\mathrm{avg}} = 4.7$ at all and can be ruled out.  The early light curve of \gwgrb{} displayed a long rise, which requires (for any structure) that $\thobs \gg \thC$.  This is clearly not possible when $\thobs \lesssim 10^\circ$, therefore necessarily $\thobs \sim 20^\circ$.   

\section{Data Fitting Methodology} \label{sec:data}

We focus here on two questions: \emph{parameter estimation}, inferring credible values of parameters in a particular model, and \emph{model comparison}: inferring which model of a set is most plausible.

Both questions involve computing or estimating the \emph{posterior} probability distribution $p({\bm{\theta}} | D, \mathcal{M}) \propto p(D | \bm{\theta}, \mathcal{M}) \times p(\bm{\theta} | \mathcal{M})$, where $p(D | \bm{\theta}, \mathcal{M})$ is the \emph{likelihood} and $p(\bm{\theta} | \mathcal{M})$ is the \emph{prior}.  Here and following, $\mathcal{M}$ will denote a particular theoretical model (and its required assumptions), ${\bm{\theta}}$ denotes the vector of free parameters within the model, and $D$ the full set of observational data (times, filters, fluxes, etc). 

A first critical step is deciding on the prior distribution to be used on the fit parameters. For most model parameters ($E_0$, $p$, $\epsB$, etc) a non-informative prior is commonly used: either uniform between upper and lower bounds (e.g. $p$) or uniform in the logarithm of the parameter if the prior is desired to be scale-invariant (e.g. $E_0$, $\epsB$).  For the viewing angle $\thobs$, the non-informative prior is the \emph{isotropic} prior which is uniform in solid angle: $p(\thobs) = \sin \thobs$ or $p(\cos \thobs) = 1$ in the domain $\thobs \in [0, \pi/2]$ \citep[used, for example, in][]{LIGOStandardSiren}.   

Given additional information, one can of course build informative priors on model parameters.  In the case of multi-messenger GW-GRB observations, one can use the posterior distribution on the GW inclination $\iota$ as a prior on the GRB viewing angle $\thobs$, assuming the GRB jet is parallel with the total angular momentum vector of the merging system \citep[first used in][]{Troja:2018GRB170817A}.  Utilizing a \ac{GW} posterior on $\iota$ as a prior on $\thobs$ naturally incorporates the \ac{GW} observational constraints, uncertainties, and degeneracies of the \ac{GW} analysis into the \ac{GRB} analysis, marginalizing over all \ac{GW} parameters.  
A simultaneous analysis of both the \ac{GRB} and \ac{GW} data with a joint model would reveal the full posterior over all parameters, an approach extensively discussed in \citep{Gianfagna:2023aa}. 

The following sections discuss additions to our parameter estimation and model comparison methodology: incorporating astrometric observations in the likelihood, incorporating Poisson statistics for low count X-ray observations, and utilizing posterior predictive densities for model comparison.

\subsection{Astrometric Data} \label{sec:vlbidata}

Astrometric observations of GRB afterglows (via radio \ac{VLBI} or high precision optical/infrared imaging, for example) can constrain the position and proper motion of the afterglow centroid as well as the size and shape of its image. Here we detail how these observations are incorporated into our full afterglow modelling pipeline, allowing for simultaneous fitting of photometric and astrometric observations.  Our approach utilizes the measured sky positions directly, an improvement over previous work which focused on the average apparent velocity between observations which will introduce a bias if they are assumed to be normally distributed \citep{Hotokezaka:2019aa, Mooley:2022vs}.

An astrometric observation of a GRB afterglow centroid results in three measurements: the two coordinates locating the centroid on the sky $x_{\RA}$ and $x_{\dec}$ and the observed flux $F_\nu$ at that location  These three measurements have uncertainties described by a covariance matrix $\Sigma^2$.

To include these observations in our afterglow modelling, we project the theoretical centroid position on to the sky.  We locate the (unknown) GRB origin at the point ($\RA_0$, $\dec_0$) and orient the jet direction with a Position Angle $\PA$.  The angular displacement of the centroid from the jet origin is $x_c$ as computed in Section \ref{sec:centroidmodel}.  The theoretical centroid position is then:
\begin{align}
    \tilde{x}_{\RA} &= \xcen \sin \PA + \RA_0 \\*
    \tilde{x}_{\dec} &= \xcen \cos{\PA} + \dec_0
\end{align}
We assume the data is described by a multivariate normal distribution over ${\bm{d}_{ast.}} = (F_\nu, \tilde{x}_{\RA}, \tilde{x}_{\dec})$ with covariance matrix $\Sigma$.  The likelihood for these observations are then:
\begin{align}
    &p(\bm{d}_{ast.} | {\bm \theta}) = \left( 2\pi\right)^{-3/2} |\Sigma|^{-1/2} \times \exp \Big[ \\*
    &\qquad -\frac{1}{2} \left(
    \bm{d}_{ast.} - \bm{d}(\bm{\theta})\right)^T \Sigma^{-1} \left(
    \bm{d}_{ast.} - \bm{d}(\bm{\theta})\right)\Big] \nonumber
\end{align}
This approach adds three model parameters to $\bm \theta$: the coordinates of the GRB origin $\RA_0$ and $\dec_0$ and the position angle $\PA$.  We take the prior on $\PA$ to be uniform in $[0^\circ, 360^\circ]$, and the priors on $\RA_0$ and $\dec_0$ to be uniform in a box $[-10 \text{ mas}, 10 \text{ mas}]$ about a reference point located within the observed sky positions.  Since $\bm{d}_{ast.}$ includes the observed flux $F_\nu$ at that epoch, this observation should not be separately included in the non-astrometric light curve data, as doing such would double-count the data point. Including these $F_\nu$ in $\bm{d}_{ast.}$ is more appropriate so potential covariances between $F_\nu$ and $\tilde{x}_{\RA}, \tilde{x}_{\dec}$ may be correctly accounted for.

It should be noted: given two position measurements on the sky with Gaussian uncertainties, the distance between the two points will follow the Hoyt distribution (the Rayleigh distribution if the Gaussians are circular) \citep{Hoyt:1947aa}.  That is, the likelihood distribution for the apparent velocity (which is proportional to the distance between the two observations) is not itself a Normal distribution, as assumed in \citet{Hotokezaka:2019aa, Mooley:2022vs}.  

In current applications to \gwgrb{} we assume $\Sigma^2$ is diagonal, and place the centroid reference point at the position of the first observation at $T_0+75$ days consistent with reported observations \citep{Mooley:2018, Ghirlanda:2019, Mooley:2022vs}.

\subsection{Poissonian Likelihoods} \label{sec:Poisson}

Late-time observations of GRB afterglows (ie. after the jet-break) can provide critical information about the jet structure and energetics and aid in breaking the viewing angle degeneracy (see, for example, \citealt{Zhang:2015aa, Troja:2016aa}). However, since afterglows decay strongly post-jet break any X-ray observations made at these times may contain only a small number of counts and be subject to Poissonian statistics.  In GRB analysis it is usually sufficient to fit X-ray light curves with a simple $\chi^2$-style likelihood, implicitly assuming or explicitly binning the data such that Gaussian statistics apply \citep[e.g.][]{Evans:2009aa, Ryan:2015aa}.

Assuming Gaussian statistics when analyzing low-count Poisson data can introduce bias \citep{Cash:1979aa, Humphrey:2009uz}. To allow for fits on low-count observations directly without requiring rebinning or risking a bias in the modeling, we have moved to a Poissonian description of our X-ray data which we review here \citep{Helene:1984aa, Gehrels:1986, Kraft:1991aa}. 

A simplified X-ray observation consists of a measured integer number of counts $N_i$ within an aperture $A_s$ and a non-integral background count $b_i$ corrected to the same aperture at time $t_i$.  The likelihood for $N_i$ given a theoretical predicted count $s$ is the Poisson distribution:
\begin{align}
    p(N_i | \bm{\theta}, b_i) &= \frac{1}{N_i!}\left(b_i + s(t_i, \bm{\theta})\right)^{N_i} \label{eq:pois} \\*
    &\qquad \times \exp\left(-b_i -s(t_i, \bm{\theta})\right) \nonumber
\end{align}
The expected background $b_i$ is typically estimated from a nearby source-free region with an aperture area $A_b$ larger than the source aperture $A_s$.  If $B_i$ counts are detected in this region the expected background in the source region is $b_i = B_i \times A_s / A_b$.  Using this estimate for $b_i$ in directly Equation \eqref{eq:pois} neglects uncertainty in the background rate.  Uncertainty may be included by treating the background as a Poisson-distributed variable:
\begin{equation}
    p(B_i | b) = \frac{1}{B_i!}\left(b \frac{A_b}{A_s}\right)^{B_i} \exp\left(-b \frac{A_b}{A_s}\right)\ . \label{eq:poisBkg}
\end{equation}
Assuming a uniform prior on $b$, its posterior distribution is then:
\begin{equation}
    p(b | B_i) = \frac{1}{B_i!} \left(\frac{A_b}{A_s}\right)^{B_i+1} b^{B_i} \exp\left(-b \frac{A_b}{A_s}\right)\ . \label{eq:poisBkgPost}
\end{equation}
Then the likelihood for $N_i$ counts in the source region becomes:
\begin{align}
    p(N_i | \bm{\theta}, B_i) &= \int db\ p(N_i | \bm{\theta}, b) p(b | B_i) \\*
        &= \frac{1}{B!} \left(\frac{A_b}{A_s}\right)^{B+1} \left(1 + \frac{A_b}{A_s}\right)^{-N-B-1} \nonumber \\*
        &\times \sum_{k=0}^N \binom{N+B-k}{B} \frac{1}{k!} \left(1 + \frac{A_b}{A_s}\right)^k s^k \nonumber \\*
        &\times \exp\left(-s\right) \label{eq:poisBkgUnc}
\end{align}
\begin{figure}
    \includegraphics[width=\columnwidth]{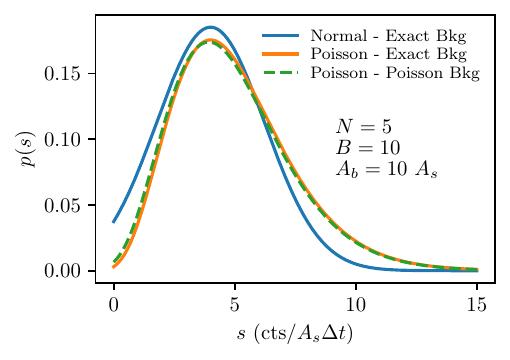}
    \caption{\label{fig:poisson} Posterior distributions on the source flux $s$ given observed counts $N=5$ and background (Bkg) counts $B=10$ under different assumptions of the data distribution. In each case the posterior $p(s) = p(s | N)$ is computed from the given likelihood using a uniform prior on $s > 0$.}
\end{figure}

Figure \ref{fig:poisson} compares these three approaches to constructing a likelihood for $N$: the Normal distribution, Poisson distribution with an exact background $b$, and Poisson distribution with an uncertain background for a mock low-count observation.  When $N, B \gg 1$ all three approaches coincide, and the details of the likelihood function become less important.  The particular choices here, $N = 5$ and $b \approx 1$, are similar to late-time observations of \gwgrb{} with \Chandra{} \citep{Troja:2022aa}.
For these parameters the uncertainty in the background rate is small and the exact-background likelihood Equation \eqref{eq:pois} provides a good approximation to the full likelihood given by Equation \eqref{eq:poisBkgUnc}.  The Normal approximation produces the correct peak position for the source flux ($s\sim 4$) but is badly skewed compared to the Poisson distributions, over-weighting the low-count-rate case relative to the high count rate case.  In applications we use the exact-background Poisson approach (Equation \eqref{eq:pois}) for its computational simplicity. 

Theoretical models typically do not directly predict the observed source count rate $s$ but rather the flux density intrinsic to the source $F_\nu$.  Formally, to compute an expected count $s$ the theoretical flux density must be forward modelled: corrected for extinction in the host galaxy and Milky Way, convolved with instrument response, and integrated over the exposure time and aperture particular for the observation performed.  In lieu of performing this entire procedure within the likelihood function, we make a simplifying assumption and compute a single flux-to-counts conversion factor $\mathtt{f2c}_i$ for each observation which we take to be fixed (see \citet{Troja:2022aa} for details of the X-ray data reduction).  This multiplicative factor converts a flux density $ F_\nu$ at frequency $\nuobs$ into an expected source count as $s = \mathtt{f2c} \times F_\nu$, and is the same approach utilized by \citet{Hajela:2022}.  The likelihood for a single observation, assuming an exact background, is then:
\begin{align}
    p(N_i | \bm{\theta}, b_i) &= \frac{1}{N_i!}\left(b_i + \mathtt{f2c}_i \times F_\nu(t_i, \nu_i, \bm{\theta})\right)^{N_i} \nonumber \\*
    & \qquad \times \exp\left(-b_i -\mathtt{f2c}_i \times F_\nu(t_i, \nu_i, \bm{\theta})\right) \label{eq:poisApprox}
\end{align}
Equation \eqref{eq:poisApprox} is an improvement over previous approaches which assume Gaussian statistics, especially when $N$ is only $\gtrsim 1$. When implemented in an MCMC parameter estimation pipeline it is equivalent to using the XSPEC C-statistic, the difference being a constant that depends only on the data \citep{Arnaud:1996aa}.  In \gwgrb{} the broadband spectrum is tightly constrained, so assuming a fixed flux-to-counts conversion (e.g. that $\mathtt{f2c}_i$ is independent of $\bm{\theta}$) introduces only small errors.  A future improvement would be to improve the forward modelling and allow the flux-to-counts conversion to depend directly on the modelled spectrum.

For faint observations $N \sim $ a few, a Poisson likelihood constrains the source flux to be brighter than a Gaussian likelihood would. Improperly applying Gaussian statistics introduces a bias that can hinder measurements of the true afterglow flux, determining the jet geometry, and identification of new late time emission sources. Table \ref{tab:Summary} shows the modest effect this has on analysis of \gwgrb{}.

\subsection{Bayesian model comparison} \label{sec:modelcomparison}

Given a set of models with plausible explanations for observed data (for instance, a lone GRB afterglow or a GRB afterglow plus a new late-time component), determining which model the data favours is a problem of \emph{model comparison}.  Model comparison is a difficult problem with many possible approaches.  We outline two general approaches below: the Bayes Factor \citep[e.g.][]{Kass:1995aa} and predictive accuracy \citep[e.g.][]{Vehtari:2012aa}. While the Bayes Factor formally answers the question ``What are the relative odds between two models given some data?'', its strong dependence on the prior (stronger than in parameter estimation) make it unsuitable for problems without a well-motivated physical prior. 

In the context of \gwgrb{}, our modelling of the potential excess in late-time X-ray observations (see Section \ref{sec:170817}) is purely phenomenological and does not have a well-motivated prior. Therefore, we rely on the predictive accuracy framework using two metrics, the Widely Applicable Information Criterion (WAIC) \citep{Watanabe:2010aa} and Leave One Out Cross Validation (LOO) \citep{Stone:1977aa, Vehtari:2015aa}.  All approaches to the model comparison problem come with their own benefits and drawbacks, we find it best to use a variety of methods and not overly rely on the results (and potential biases) of any one technique.  In the rest of this section we review these approaches in order.

\subsubsection{Bayes Factors}

The relative odds between two models $\mathcal{M}_1$ and $\mathcal{M}_2$ given data $D$ follow from Bayes' theorem and are given by
\begin{equation}
    \frac{p(\mathcal{M}_1 | D)}{p(\mathcal{M}_2 | D)} = \frac{p(D|\mathcal{M}_1) \times p(\mathcal{M}_1)}{p(D|\mathcal{M}_2) \times p(\mathcal{M}_2)} \equiv \frac{\mathcal{Z}_1}{\mathcal{Z}_2 } \times \frac{p(\mathcal{M}_1)}{ p(\mathcal{M}_2)}.
\end{equation}
The ratio $\mathcal{Z}_1 / \mathcal{Z}_2$ is the Bayes factor, and would be all that mattered for the purpose of comparing models if the initial probabilities of both models $p(\mathcal{M})$ were set to be equal. A first subtlety thus is that there may well be a priori reasons to value one model over another and where $p(\mathcal{M}_1)/p(\mathcal{M}_2)$ is not set to unity\footnote{In the case of GRB afterglows, \cite{SarinLaskyAshton:2019} provide an example of one such situation, where they compare different central engine models (magnetar versus black hole), but take the a priori ratio $p(\mathcal{M}_1)/p(\mathcal{M}_2)$ to be given by externally assessed  odds that a general neutron star collapse produces a magnetar.}.

The other subtlety is that the evidence $\mathcal{Z}$ is defined as the average probability of obtaining a given data set from a model across the parameter space of the model using the prior as a weight function, that is
\begin{equation}
\mathcal{Z} = p(D|\mathcal{M}) = \int p(D|\bm{\theta}, \mathcal{M}) p( \bm{\theta} | \mathcal{M} ) d \theta,
\end{equation}
which therefore renders the evidence itself highly sensitive to the prior. Because the prior is normalized, the evidence favours more compact models: whenever a model parameter space includes a range for its parameters that contributes unlikely solutions, this dilutes the evidence for that model even if the low-likelihood region is `only' due to a broad domain choice for an otherwise meaningful parameter. Model selection in particular therefore stresses the notion that there is no such thing as an uninformative prior. 

We note that in a sense the Bayes Factor is \emph{more} sensitive to the choice of prior than simple parameter estimation.  Suppose an experiment with data $D$ and the prior of some parameter $\theta$ is chosen to be uniform in a range $[\theta_0 , \theta_0 + \Delta \theta]$.  If the data then provides a strong constraint on $\theta$, say $\theta \sim \theta^* \pm \sigma$, then the posterior will be independent of the prior so long as it is large enough to contain the region favoured by the likelihood: $p(\theta | D)$ will be independent of $\Delta \theta$ so long as $\Delta \theta \gg \sigma$.  The evidence, however, will be approximately $Z \sim p(D|\theta^*) \sigma / \Delta \Theta$: inversely proportional to $\Delta \theta$ no matter its value.  

This is, of course, correct and a feature of the Bayes Factor approach to model comparison: if two models fit the data equally well, the one which made the more specific prediction (admits a smaller region of parameter space) should be preferred \citep{Jaynes:2003aa}.  In this work, however, this presents a problem. Our model of the late time X-ray excess is purely phenomenological.  We do not know the full range of possible processes that could produce X-ray emission in this system, and so putting a meaningful prior on the model parameters is very difficult.  The usual approach of simply admitting wide bounds on the parameters directly biases the result.  For this reason we prefer not to use Bayes factors in this application, and consider predictive accuracy techniques.

\subsubsection{Predictive Accuracy}

An alternative method of comparing models is by their \emph{predictive accuracy}.  Instead of answering the question ``what are the odds in favour of $\mathcal{M}_1$ over $\mathcal{M}_2$?" predictive methods answer the question ``which model is expected to give more accurate predictions for new data?''  This framing allows predictive models to sidestep questions of a model's physicality or interpretation by evaluating performance solely in the data-space.  This is a benefit when working with phenomenological models with uncertain or ill-posed priors, such as the existence of a new X-ray component in \gwgrb{}, which are not suitable for comparison via Bayes Factors.  A further advantage is their relative computational simplicity: predictive measures require only MCMC samples of the posterior distribution, generated by the investigator's method of choice.

Given a fit of data $D$ to a model $\MM$, the probability of observing a new datum $d'$ is given by the predictive density $p(d'|D,\MM) = \int p(d'|\theta,\MM) p(\theta | D,\MM) d\theta$, where $p(d'|\theta,\MM)$ is the model likelihood and $p(\theta | D,\MM)$ is the posterior from the fit to $D$.  A common measure of predictive accuracy is the logarithm of the predictive density averaged over all possible data sets $D' = \{ {d_i}'\}$ generated by the true model, the \emph{expected log pointwise predictive density} (elpd) \citep{Gelman:2013aa}:
\begin{equation}
    \mathrm{elpd} \equiv \sum_i \langle \log p ({d_i}' | D, \mathcal{M})\rangle_{\mathrm{true}} \ . 
\end{equation}
A model with high elpd produces data from the true model with high probability, and should be favoured over one which does not. Of course, the elpd is impossible to evaluate in practice given that the true model is unknown.  Predictive measures for model comparison approximate the elpd using the known data only.  We make use of two measures, leave-one-out cross-validation (LOO) \citep[e.g.][]{Stone:1977aa} and the Widely Applicable Information Criterion (WAIC) \citep{Watanabe:2010aa}.

The LOO approximates the elpd using the predictive density for $\MM$ fit to $D_{-i} = D 
\setminus d_i$, the data set with the $i$'th data point left out.  That is, LOO supposes $ \langle p ({d_i}' | D, \mathcal{M})\rangle_{\mathrm{true}} \approx p(d_i | D_{-i}, \MM)$ and writes:
\begin{equation}
    \mathrm{LOO}_\mathrm{elpd} = \sum_i \log  p ({d_i} | D_{-i}, \mathcal{M}) \ . 
\end{equation}
The leave-one-out predictive density $p ({d_i} | D_{-i}) = \int p(d_i | \theta) p(\theta | D_{-i}) d\theta$ can be calculated by importance sampling draws from the posterior $\theta_j \sim p(\theta | D)$ from the fit to the full data set with weights $w_j \approx 1/p(d_i | \theta_j)$.  In practice using $1/p(d_i | \theta_j)$ exactly leads to an unstable computation. We use the Pareto-Smoothing procedure of \citet{Vehtari:2015aa, Vehtari:2015wb} to stabilize the weights and compute $p ({d_i} | D_{-i})$.

The WAIC is a similar approximant which is equivalent to the LOO for large datasets \citep{Watanabe:2010aa, Gelman:2013aa}.  It is computed by evaluating the log predictive density directly on the known data $D$ and subtracting a measure of the effective number of parameters $p_{\mathrm{WAIC}}$ to correct for over-fitting.
\begin{align}
    \mathrm{WAIC}_\mathrm{elpd} &= \sum_i \log  p (d_i | D, \mathcal{M}) - p_{\mathrm{WAIC}}\\*
    p_{\mathrm{WAIC}} &= \sum_i \mathrm{var}_{\mathrm{post}, \MM} \log  p (d_i | \theta)
\end{align}
Both terms may be computed directly from $N$ posterior samples $p (d_i | D) = \sum_j p(d_i | \theta_j) / N$ and $\mathrm{var}_{\mathrm{post}, \MM} \log  p (d_i | \theta) = \sum_j \left[ \log  p (d_i | \theta_j) - \langle \log  p (d_i | \theta) \rangle\right]^2/(N-1)$.

Models can be compared via their LOO or WAIC scores, a model with a higher score is likely to have better predictive accuracy and may be preferred.  The significance of a difference in predictive score $X$ (either LOO or WAIC) can be computed from the variance of the differences over the data set.  Following \citet{Vehtari:2015aa}, since each score is a sum over contributions from each data point, $X = \sum_i X_i$, the standard error can be obtained:
\begin{equation}
    \sigma_{\mathrm{SE}}(\Delta X) = \sqrt{\frac{N_D}{N_D-1}\sum_i^{N_D} \left[{\Delta X_i - \frac{1}{N_D}\Delta X}\right]^2} \label{eq:predErr}
\end{equation}
For sufficiently large data sets and distinct models, the distribution of $\Delta X_i$ is almost normal and Equation \eqref{eq:predErr} is an accurate estimate of the standard error in $\Delta X$ \citep{Vehtari:2015aa, Sivula:2020tb}.

The WAIC and LOO provide two ways of approximating the elpd, measuring predictive accuracy of a model.  This approach does \emph{not} take into account the physicality or prior probability of the models: it will prefer a polynomial fit to the data over a sophisticated simulation if the polynomial is deemed more predictive.  All computations are performed as posterior estimates, so they inherit identical dependence on the prior as the parameter estimation itself.  Models with too many parameters are appropriately penalized, as such models tend to have large variability in their predictions outside the narrow range of observed data.  The WAIC estimate penalizes this explicitly with the $p_{\mathrm{WAIC}}$ parameter, the LOO estimate implicitly by removing each data point's contribution to the posterior before computing its predictive density.

While both the WAIC and LOO are non-biased estimators of the elpd in the limit of large datasets, in practice (with only finite data) they can produce differing estimates of the elpd and its uncertainty.  We prefer to compute both, and in the absence of specific reasons to trust one metric over the other, favour the more conservative result of the two.

\section{Application to GW170817} \label{sec:170817}

The landmark \gwgrb{} was the first EM counterpart to a GW event, confirmed the binary neutron star mergers as the progenitors of short GRBs, confirmed the existence of the kilonova transient, and prompted a re-evaluation of GRB afterglow theory.  Three years after the GRB, X-ray observations began to show an excess above predicted forward-shock jet models \citep{Troja:2020aa}.  Attempts to fit the light curve required large opening and viewing angles, in tension with the observed superluminal apparent velocity.  This excess has continued \citep{Troja:2022aa}, raising questions as to whether a new emitting component, such as the non-thermal afterglow of the kilonova or an accreting black hole remnant, may be present \citep{Hajela:2022, Balasubramanian:2022}.

To best determine the parameters of \gwgrb{} and evaluate the potential presence of a new component, we have fit our full model to the \gwgrb{} non-thermal afterglow. In addition to our forward-shock jet model used in previous works \citep[e.g.][]{Troja:2020aa} we include the astrometric centroid position measurements, Deep-Newtonian treatment of the late-time synchrotron radiation, and Poisson likelihoods on the \Chandra{} X-ray data.  The dataset is exactly that of \citet{Troja:2022aa}, plus the additional VLBI data from \citet{Mooley:2018, Ghirlanda:2019}, the HST centroid position from \citet{Mooley:2022vs}, and two \Chandra{} observations taken at $\approx 1634$ days and $\approx 2012$ days after the burst (PI: Troja).  

All parameter estimation was done via MCMC sampling of the posterior, performed with the Parallel Tempering sampler in \emcee{} \texttt{v2.2.1} \citep{Earl:2005aa, Foreman-Mackey:2013aa}.  We utilized 20 temperature levels geometrically spaced between $1$ and $10^6$, $100$ walkers per level, and ran for $10000$ iterations discarding the first $2500$ as burn-in.  The retained $7500$ steps of $100$ walkers each typically show an auto-correlation time of $\approx 10$, yielding $\sim 75000$ independent posterior samples for each fit.  Afterglow emission and centroid sky position were computed with \afterglowpy{} \texttt{v0.8.0}.

For our prior on $\thobs$ we utilize LIGO and Virgo published posterior distributions on the inclination of \gwgrb{} under different assumptions about cosmological parameters \citep{LIGOStandardSiren}. Taking the inclination to be equal to viewing angle, the distribution assuming the \Planck{} value of $H_0$ is well fit by the Gaussian $p(\cos \thobs) \propto \mathcal{N}(\mu=0.985, \sigma=0.070)$, with the additional restriction $0 \leq \cos \thobs \leq 1$.  We take this as our ``GW prior'' on $\thobs$ for \gwgrb{}.  Small deviations from the true GW posterior distribution induced by the Gaussian approximation distort the error estimation only in the region above 3 sigma, emphasizing the importance of  properly reproducing the posterior distribution \citep{Gianfagna:2023aa}. 

We first consider the full model, then examine the contributions of the individual new features.

\subsection{Fit of the full model}

\begin{figure*}
    \begin{center}
    \begin{tabular}{cc}
    \includegraphics[width=0.5\textwidth]{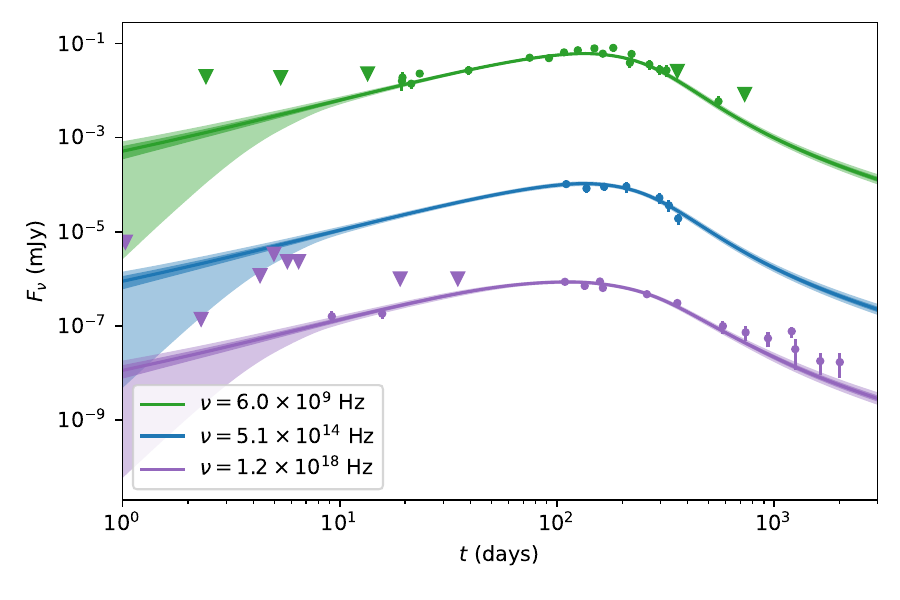} & \includegraphics[width=0.5\textwidth]{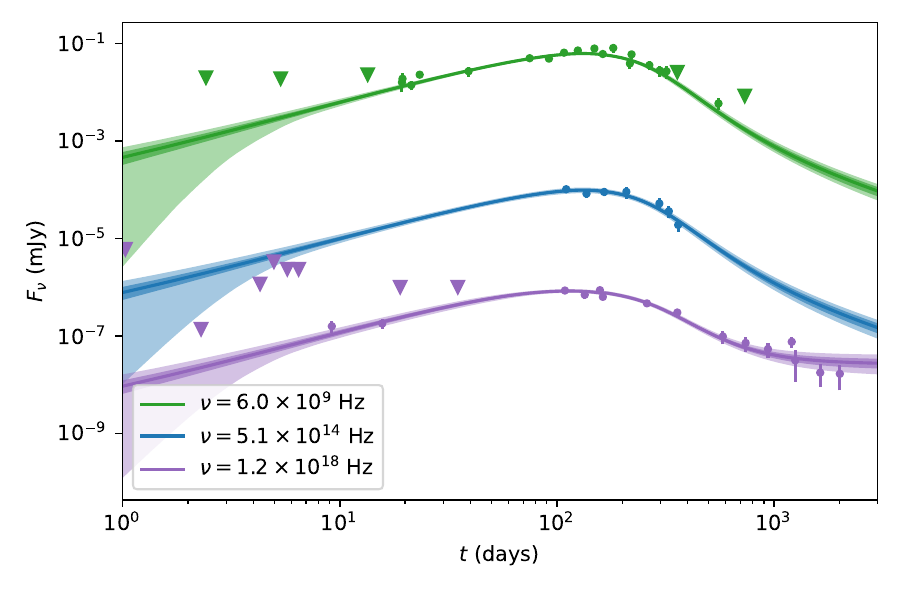}    
    \end{tabular}
    \end{center}
    \caption{\label{fig:fitLC} The \gwgrb{} light multiband light curve and fits with the full model.  Radio, optical, and X-ray data in green, blue, and purple respectively.  Triangles show 3 $\sigma$ upper limits, only data points within 20\% of the target frequency are shown.  Solid bands show the modelled afterglow emission after MCMC fitting. The solid curve shows the median flux as a function of time, the pale bands show the 68\% and 95\% quantiles of the flux at each time.  Left panel: fit with a Guassian jet only.  Right panel: fit with a Gaussian jet plus additional late-time constant luminous X-ray component $L_X$.}
\end{figure*}

We considered both Gaussian and Power-Law (Equations \eqref{eq:GaussianJet} and \eqref{eq:PowerlawJet}) structured jets. For each jet model we performed two fits: with and without an additional constant-in-time luminous X-ray component $L_X$. We choose the new component to be constant for two reasons: it requires only a single extra parameter and so will not be overly-penalized in the model comparison, and the observed X-ray flux does not currently show an obvious rise with time, so any new component must be only slowly evolving.  We note this ``model'' is purely phenomenological, it deliberately is meant to only capture average behaviour and is agnostic to specific physical models.

The additional component is a simple additive contribution to the X-ray flux $F_\nu = F_{\nu, \mathrm{afterglow}} + F_{\nu, X}$.  We parameterize the component in terms of its isotropic luminosity $L_{X, 38} = L_X / 10^{38}$ erg s$^{-1}$, computed over the 0.3-10.0 keV X-ray band assuming a $\beta = -0.585$ power-law spectrum.  Since our likelihood function for the X-ray data ignores spectral variations, the net effect of adding Equation to our model is to add a single constant flux with strength $F_{\nu, X} \propto L_{X,38}$.  

When fitting for the centroid motion we include the origin point location \refdec{}, \refra{}, and position angle PA as fit parameters.  The full parameter set is:
\begin{align}
    \Theta &= \left\{ \thobs, E_0, \thC, \thW, b, n_0, p, \epse, \epsB, \xiN, \right.\nonumber \\*
    &\qquad \left . \refdec, \refra, \posang, \LX \right \}\ ,
\end{align}
where $b$ is only included in the Power-Law jet\footnote{The $b$ here denotes the asymptotic power-law index of the jet structure function $E(\theta)$ and should not be confused with the background count rate in Section \ref{sec:Poisson}.} and \LX{} is only included if an extra luminosity is being considered.

\begin{figure}
    \begin{center}
    \begin{tabular}{c}
    \includegraphics[trim=0 3.5cm 0 3.5cm, clip, width=0.98\columnwidth]{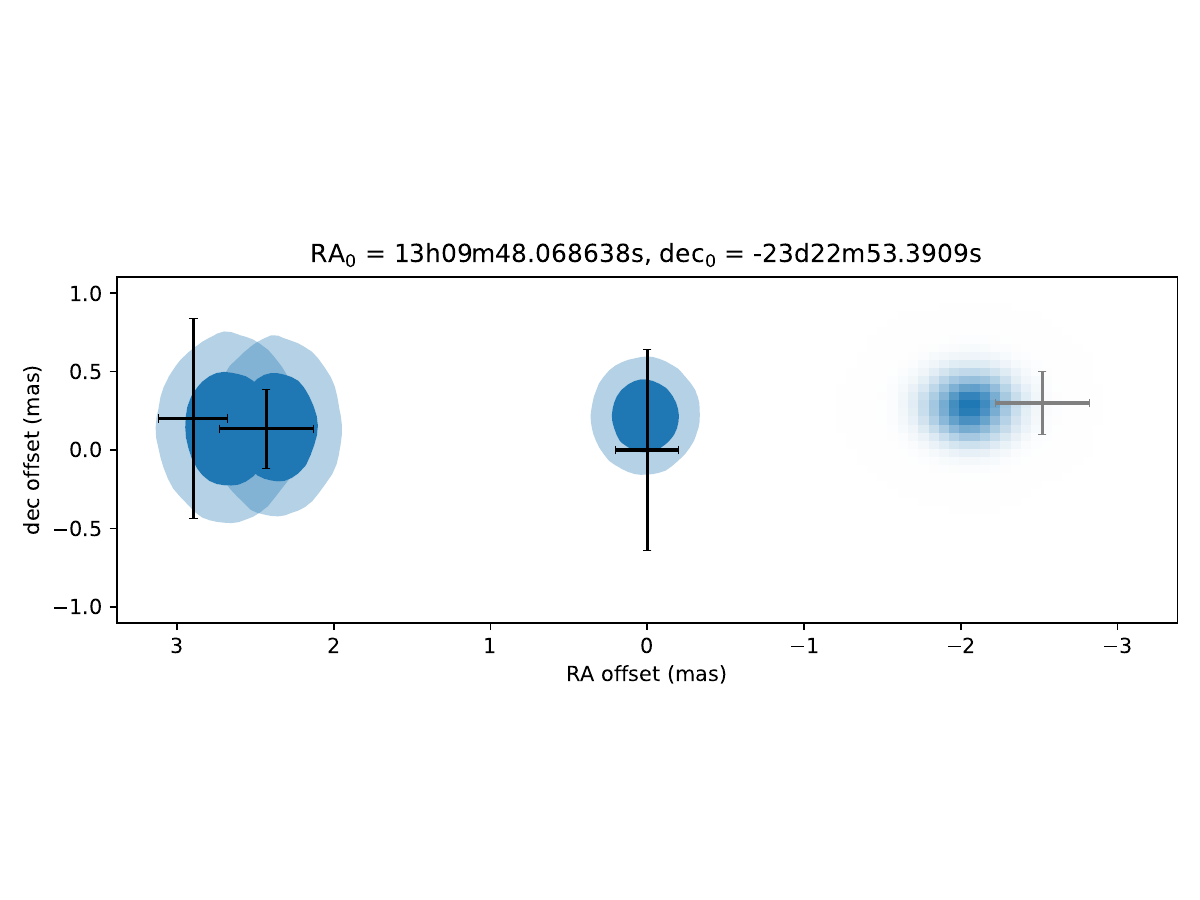} \\
    \includegraphics[trim=0 3.5cm 0 4.5cm, clip, width=0.98\columnwidth]{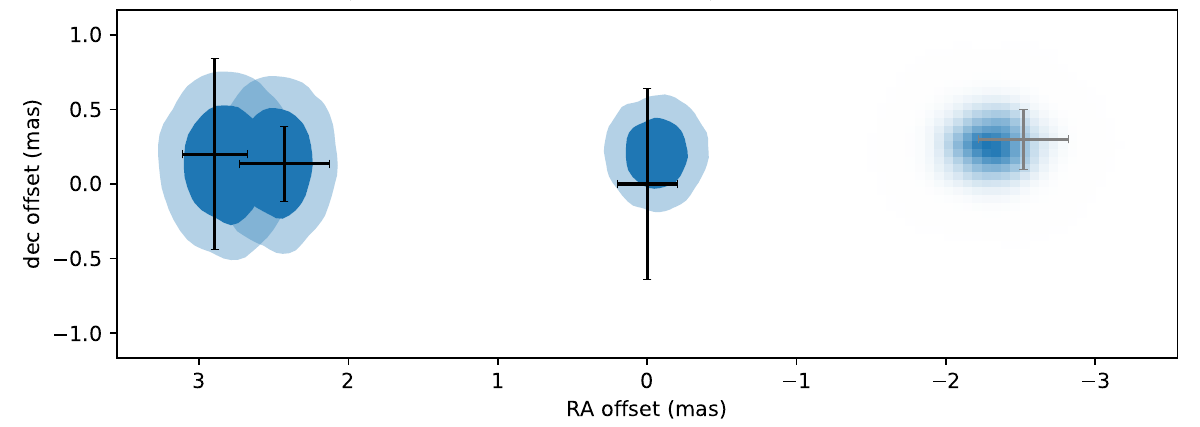}    
    \end{tabular}
    \end{center}
    \caption{\label{fig:fitPosition} The position of the \gwgrb{} centroid on the sky and fits with the full model.  Black data points are radio VLBI measurements \citep{Mooley:2018, Ghirlanda:2019}. The grey data point is the \HST{} kilonova position, assumed to be the burst origin \citep{Mooley:2022vs}.  Blue regions show the 68\% (dark) and 95\% (light) posteriors of the fit location at each epoch.  Pale blue cloud on the right side of each panel shows the posterior distribution on the burst origin.  Top panel: fit with a Gaussian structured jet only.  Bottom panel: fit with a Gaussian structured jet plus a constant additional X-ray component.}
\end{figure}

\begin{deluxetable*}{CCCCCC}
    \tablecaption{Fits with the full model.  Top section shows fit parameters, empty entries were not relevant to that model.  Middle section shows quantities derived from the fit parameters: the total energy of the afterglow jet and the ratio of viewing angle to opening angle.  Bottom section shows model comparison quantities: the elpd approximant, effective number of parameters, and difference with highest scoring model.  Uncertainties are all given at 68\%.  Models are: \texttt{GJ} - Gaussian jet only,  \texttt{GJ+LX} -  Gaussian jet with late-time $L_X$,  \texttt{PLJ} - Power-law jet only, and \texttt{PLJ+LX} - Power-Law Jet with late-time $L_X$.   \label{tab:fullResults}}
    \tablehead{\colhead{Parameter} & \colhead{Unit} & \colhead{\texttt{GJ}} & \colhead{\texttt{GJ+LX}} & \colhead{\texttt{PLJ}} & \colhead{\texttt{PLJ+LX}}}
\startdata
\theta_{\mathrm{obs}} & \text{deg} & {20.8}_{-1.4}^{+1.7} & {19.2}_{-1.2}^{+1.4} & {21.5}_{-1.5}^{+1.8} & {19.4}_{-1.2}^{+1.5}\\
\log_{10}E_0 & \text{erg} & {54.35}_{-0.63}^{+0.41} & {54.0}_{-1.0}^{+0.6} & {54.50}_{-0.54}^{+0.36} & {54.3}_{-1.2}^{+0.5}\\
\theta_{\mathrm{c}} & \text{deg} & {3.50}_{-0.25}^{+0.29} & {3.20}_{-0.22}^{+0.25} & {2.44}_{-0.21}^{+0.25} & {2.09}_{-0.18}^{+0.21}\\
\theta_{\mathrm{w}} & \text{deg} & {25}_{-10}^{+11} & {23.7}_{-9.3}^{+9.9} & {11.7}_{-1.1}^{+1.6} & {11.0}_{-1.1}^{+2.2}\\
b &  &   &   & {9.41}_{-0.79}^{+0.43} & {9.28}_{-0.94}^{+0.53}\\
\log_{10}n_0 & \text{cm}^{-3} & {-1.75}_{-0.62}^{+0.46} & {-2.31}_{-0.98}^{+0.64} & {-1.65}_{-0.56}^{+0.43} & {-2.1}_{-1.2}^{+0.5}\\
p &  & {2.1223}_{-0.0093}^{+0.0095} & {2.137}_{-0.011}^{+0.011} & {2.1193}_{-0.0093}^{+0.0094} & {2.138}_{-0.011}^{+0.011}\\
\log_{10}\epsilon_e &  & {-3.42}_{-0.41}^{+0.61} & {-3.10}_{-0.62}^{+0.97} & {-3.30}_{-0.36}^{+0.53} & {-3.1}_{-0.5}^{+1.1}\\
\log_{10}\epsilon_B &  & {-4.02}_{-0.43}^{+0.62} & {-3.63}_{-0.63}^{+0.98} & {-4.04}_{-0.38}^{+0.54} & {-3.8}_{-0.5}^{+1.2}\\
\log_{10}\xi_N &  & {-0.47}_{-0.57}^{+0.36} & {-0.67}_{-0.93}^{+0.52} & {-0.31}_{-0.48}^{+0.27} & {-0.34}_{-0.77}^{+0.31}\\
\text{RA}_0 & \text{mas} & {-2.05}_{-0.25}^{+0.26} & {-2.30}_{-0.25}^{+0.25} & {-1.93}_{-0.26}^{+0.25} & {-2.22}_{-0.24}^{+0.25}\\
\text{dec}_0 & \text{mas} & {0.27}_{-0.21}^{+0.20} & {0.27}_{-0.22}^{+0.20} & {0.26}_{-0.21}^{+0.20} & {0.27}_{-0.22}^{+0.20}\\
\text{PA} & \text{deg} & {91.6}_{-3.8}^{+3.8} & {91.7}_{-3.5}^{+3.6} & {91.5}_{-3.9}^{+3.9} & {91.6}_{-3.7}^{+3.7}\\
L_{X,38} & 10^{38}\text{erg} &   & {1.48}_{-0.36}^{+0.39} &   & {1.66}_{-0.38}^{+0.41}\\
\hline
\log_{10}E_{tot} & \text{erg} & {51.92}_{-0.61}^{+0.41} & {51.5}_{-1.0}^{+0.6} & {51.85}_{-0.53}^{+0.35} & {51.6}_{-1.2}^{+0.5}\\
\theta_{\mathrm{obs}}/\theta_{\mathrm{c}} &  & {5.94}_{-0.15}^{+0.16} & {6.01}_{-0.17}^{+0.17} & {8.81}_{-0.50}^{+0.53} & {9.25}_{-0.51}^{+0.63}\\
\hline
\mathrm{WAIC}_\mathrm{elpd} &   & {432}\pm{48} & {444}\pm{46} & {422}\pm{48} & {440}\pm{46}\\
p_\mathrm{WAIC} &   & 10.3 & 9.2 & 17.1 & 11.7\\
\Delta \mathrm{WAIC}_\mathrm{elpd} &   & {-11.9}\pm{5.6}\ (-2.1\sigma) &  \texttt{best} & {-22.4}\pm{7.9}\ (-2.8\sigma) & {-4.2}\pm{2.8}\ (-1.5\sigma)\\
\mathrm{LOO}_\mathrm{elpd} &   & {431}\pm{97} & {443}\pm{93} & {418}\pm{99} & {436}\pm{94}\\
p_\mathrm{LOO} &   & 12.1 & 10.2 & 21.5 & 15.7\\
\Delta \mathrm{LOO}_\mathrm{elpd} &   & {-13}\pm{11}\ (-1.2\sigma) & \texttt{best}  & {-26}\pm{14}\ (-1.8\sigma) & {-7.1}\pm{5.8}\ (-1.2\sigma)\\
\hline
\enddata

\end{deluxetable*}

The results of the fits are summarized in Table \ref{tab:fullResults}.  The addition of the centroid position data substantially constrains the viewing angle: all models agree $\thobs \sim 20^\circ$.  Individual fits give median values of $\thobs$ in the range $19.2^\circ-21.5^\circ$, with uncertainties of $\sim 1.5^\circ$.  Fits with the additional $L_X$ show smaller viewing angles than jet-only models.  

The tight measurement of the viewing angle translates into a similarly constraint on the opening angle $\thC$: the Gaussian jets find $\thC \approx 3.5^\circ, 3.2^\circ$ while the power-law jets find $\thC \approx 2.44^\circ, 2.09^\circ$ for fits without and with additional $L_X$, respectively. Power-law jets prefer a large power-law index $b \sim 9$.  

Jet-only models require somewhat brighter late-time light curves to match X-ray observations, pushing $\thC$ to larger values.  However, this must be compensated by a proportional increase in $\thobs$ to keep the early light curve slope in agreement with observations.  This larger $\thobs$ comes into tension with large apparent superluminal velocity from the centroid observations, which shrinks as the viewing angle grows.  The centroid observations provide more constraining power than the late X-ray, limiting the range of $\thobs$ and $\thC$ between different models.

The isotropic energy and density have large uncertainties (up to an order of magnitude at the 68\% level) due to parameter degeneracies inherent in any afterglow model, but are broadly consistent with expectations. The total kinetic energy has substantial weight below $10^{52}$ erg (the rough magnetar limit) for all models. The external density is $n_0 \sim 10^{-2}$ cm$^{-3}$ in all fits, consistent with expectations for a binary neutron star merger occurring far from the center of its host galaxy and with a large amount of support within the $n_0 < 9.6 \times 10^{-3}$ cm$^{-3}$ bound measured by \citet{Hajela:2019aa} from diffuse X-ray emission.

The electron power-law index $p$ is approximately $2.12 \pm 0.01$ and $2.14 \pm 0.01$ in fits with and without $L_X$ respectively.  This is modestly smaller than the $2.17 \pm 0.01$ found in analyses from the first year of data \citep[e.g.][]{Troja:2019ab}, largely due to the shallower than expected late-time decay.  The synchrotron microphysical parameters $\epse$ and $\epsB$ are small ($\sim 10^{-3}$ and $\sim 10^{-4}$ respectively) in all fits.  The lower-bound of these posteriors are enforced by the prior to be $10^{-4}$ and $10^{-5}$.  The participation fraction $\xiN \sim 0.5$ is large in all fits, the parameter degeneracy that would allow it to take smaller values would require $\epse$ and $\epsB$ to shrink as well, which is prevented by the prior.  

The origin of the burst on the sky is found to be approximately $2.0\pm 0.2$ mas west and $2.3\pm 0.2$ mas west of the reference point, $0.5$ mas and $0.2$ mas from the kilonova location as measured by \emph{HST}, for fits without and with $L_X$ respectively \citep{Mooley:2022vs}.  The addition $L_X$ gives the model freedom to shrink the viewing angle and better reproduce the proper motion.

Overall, the Gaussian Jet model with an additional late time X-ray luminosity is preferred, although the preference is marginal: $1\sigma$ (LOO) - $2\sigma$ (WAIC) over the Gaussian Jet without an additional component. We do not find this to be convincing evidence for a new component in GW170817.

\begin{deluxetable*}{CCCCCCCCCC}
    \tablecaption{Additional fits to \gwgrb{} testing dependence on features of the modelling pipeline:  GW priors on the opening angle $\thobs$ (GW prior), proper motion of the centroid (Centroid), Deep-Newtonian emission regime (DN), and Poisson X-ray likelihoods (Poisson).  Each feature is ran with two emission models: a Gaussian jet only (\texttt{GJ}) and a Gaussian jet with additional late-time X-ray (\texttt{GJ+LX}).  Marginalized posterior estimates of the viewing angle $\thobs$ and opening angle $\thC$ (both in degrees) and the X-Ray luminosity $L_{X,38}$ (in $10^{38}$ erg/s) are given.  The final columns show whether the model with or without $L_X$ was preferred, and the difference and significance of the preference via the WAIC and LOO scores.
    \label{tab:Summary}}
    \tablehead{\colhead{Model} & \colhead{GW Prior} & \colhead{Centroid} & \colhead{DN}& \colhead{Poisson}  & \colhead{$\theta_{\mathrm{obs}}$} & \colhead{$\theta_{\mathrm{c}}$} & \colhead{$L_{X,38}$} & \colhead{$\Delta \mathrm{WAIC}_\mathrm{elpd}$} & \colhead{$\Delta \mathrm{LOO}_\mathrm{elpd}$}}
\startdata
\texttt{GJ} & \checkmark & \checkmark & \checkmark & \checkmark & {20.8}_{-1.4}^{+1.7} & {3.50}_{-0.25}^{+0.29} &   & {-11.9}\pm{5.6}\ (-2.1\sigma) & {-13}\pm{11}\ (-1.2\sigma)\\
\texttt{GJ+LX} & \checkmark & \checkmark & \checkmark & \checkmark & {19.2}_{-1.2}^{+1.4} & {3.20}_{-0.22}^{+0.25} & {1.48}_{-0.36}^{+0.39} &  \texttt{best} & \texttt{best} \\
\hline
\texttt{GJ} & &  & \checkmark & \checkmark & {42.5}_{-4.1}^{+2.4} & {6.90}_{-0.67}^{+0.44} &   & {-0.4}\pm{1.8}\ (-0.2\sigma) & {-0.4}\pm{3.4}\ (-0.1\sigma)\\
\texttt{GJ+LX} &  &  & \checkmark & \checkmark & {37.4}_{-7.5}^{+5.8} & {6.1}_{-1.2}^{+1.0} & {0.74}_{-0.33}^{+0.35} & \texttt{best}  &\texttt{best}  \\
\hline
\texttt{GJ} & \checkmark &  & \checkmark & \checkmark & {35.6}_{-4.0}^{+3.9} & {5.81}_{-0.64}^{+0.64} &   & {-1.4}\pm{2.5}\ (-0.6\sigma) & {-1.4}\pm{4.7}\ (-0.3\sigma)\\
\texttt{GJ+LX} & \checkmark &  & \checkmark & \checkmark & {27.3}_{-5.5}^{+5.5} & {4.50}_{-0.90}^{+0.90} & {1.21}_{-0.41}^{+0.42} &  \texttt{best} & \texttt{best} \\
\hline
\texttt{GJ} &  & \checkmark & \checkmark & \checkmark & {20.9}_{-1.4}^{+1.7} & {3.53}_{-0.25}^{+0.30} &   & {-12.2}\pm{5.6}\ (-2.2\sigma) & {-13}\pm{11}\ (-1.2\sigma)\\
\texttt{GJ+LX} &  & \checkmark & \checkmark & \checkmark & {19.3}_{-1.2}^{+1.5} & {3.21}_{-0.22}^{+0.26} & {1.48}_{-0.37}^{+0.38} &  \texttt{best} & \texttt{best} \\
\hline
\texttt{GJ} & \checkmark & \checkmark &  & \checkmark & {21.2}_{-1.5}^{+1.7} & {3.59}_{-0.26}^{+0.31} &   & {-16.8}\pm{6.9}\ (-2.4\sigma) & {-18}\pm{13}\ (-1.3\sigma)\\
\texttt{GJ+LX} & \checkmark & \checkmark &  & \checkmark & {19.4}_{-1.2}^{+1.4} & {3.23}_{-0.22}^{+0.25} & {1.58}_{-0.35}^{+0.38} & \texttt{best}  & \texttt{best} \\
\hline
\texttt{GJ} & \checkmark & \checkmark & \checkmark &  & {19.8}_{-1.3}^{+1.5} & {3.30}_{-0.23}^{+0.27} &   & {-6.2}\pm{3.0}\ (-2.0\sigma) & {-6.2}\pm{6.1}\ (-1.0\sigma)\\
\texttt{GJ+LX} & \checkmark & \checkmark & \checkmark &  & {19.2}_{-1.2}^{+1.4} & {3.19}_{-0.22}^{+0.25} & {1.20}_{-0.31}^{+0.32} & \texttt{best}  & \texttt{best} \\
\enddata

\end{deluxetable*}

\subsection{What does the centroid do?}

\begin{figure*}
    \begin{centering}
        \includegraphics[width=0.5\textwidth]{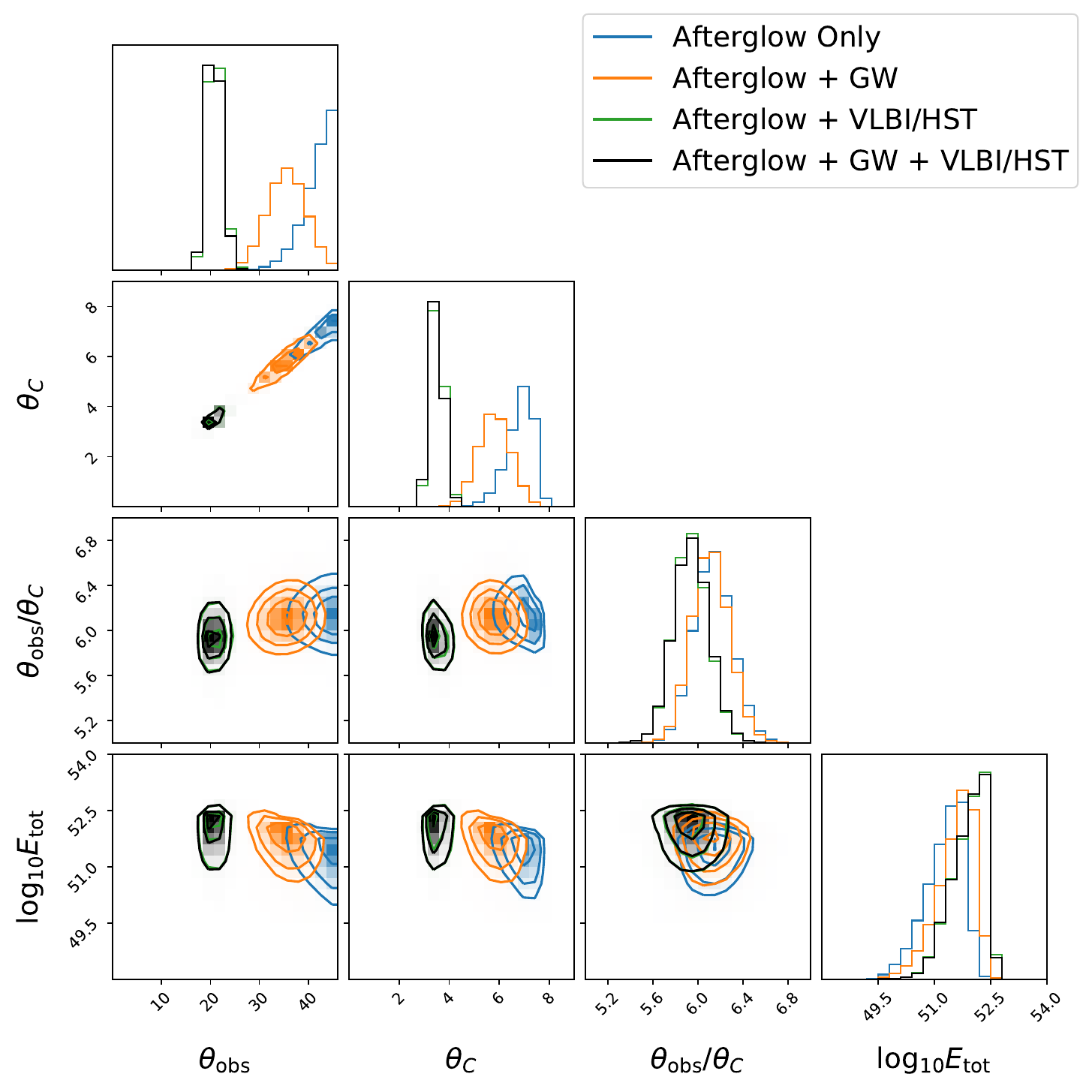}
        \includegraphics[width=0.5\textwidth]{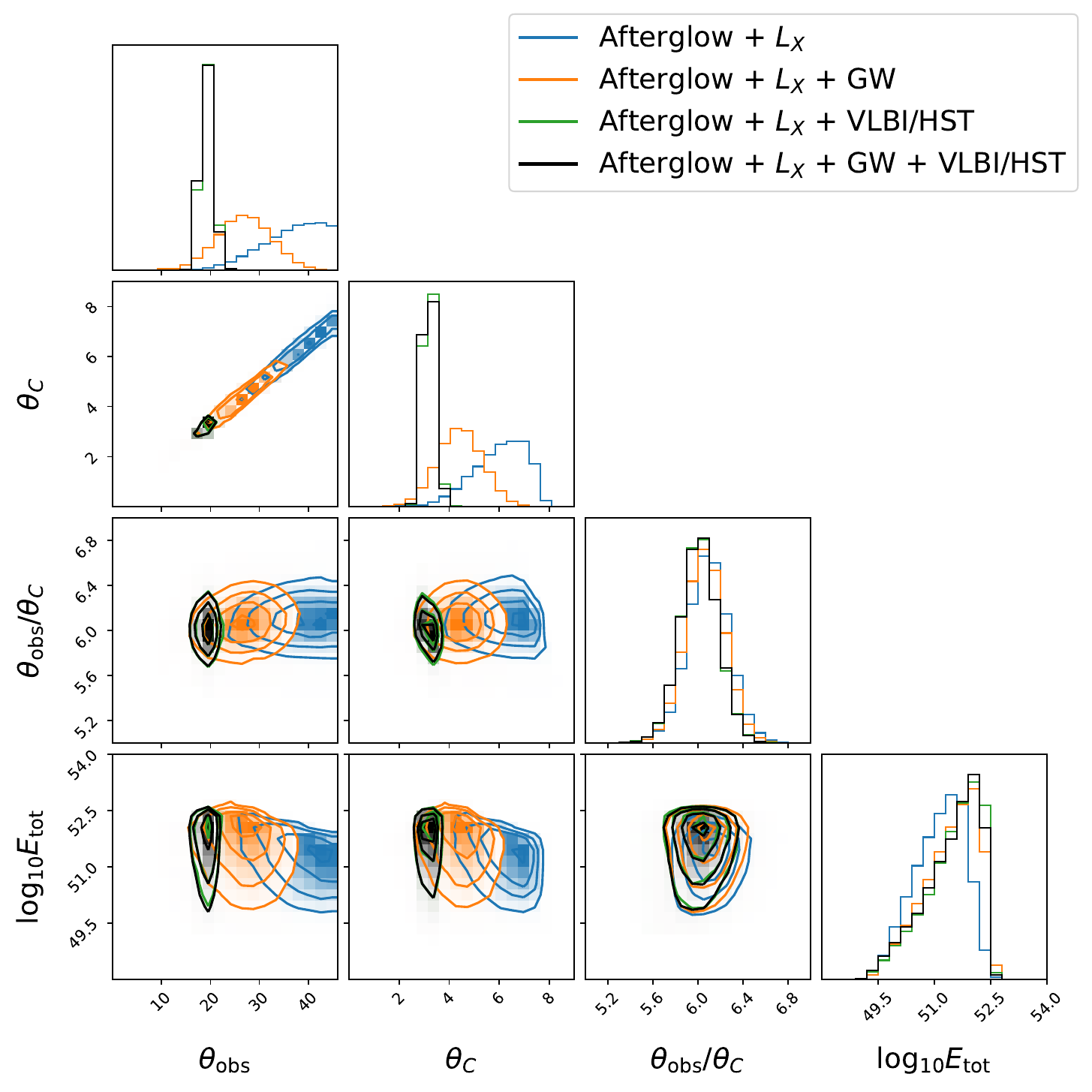}
    \end{centering}
    \caption{\label{fig:geometryInference} Projections of the posterior distribution to full data set of GW170817 for models without (left panel) and with (right panel) an extra $L_X$ component.  Each plot shows the results under four different assumptions: only fitting the afterglow light curve data (blue), including the gravitational wave prior on the viewing angle $\thobs$ (orange), including the VLBI centroid measurement (green), and including both VLBI and GW prior (black).}
\end{figure*}

Apparent superluminal motion of the afterglow radio centroid is a valuable constraint on the afterglow geometry independent of the light curve.  In Figure \ref{fig:geometryInference} we show the constraints on $\thobs$, $\thC$, $\thobs / \thC$, and $E_{\rm tot}$ when including different sets of observations for \gwgrb{}.  The ratio $\thobs / \thC$ (for our Gaussian jet model) is constrained very well in all cases.  This ratio sets the initial pre-jet break slope of the light curve through the structured jet closure relations \cite{Ryan:2020aa}.  However, absent other constraints this ratio is the only geometrical parameter well determined by the data light curve data.   

The individual angles $\thobs$ and $\thC$ are not well constrained by the light curve alone.  In the Afterglow Only fit $\thC$ is required to be large to produce a brighter Sedov phase and make up the late time X-ray excess.  This drives the viewing angle to anomalously large values $\thobs \sim 45^\circ$.  In the fit with additional $L_X$ the Sedov phase is unconstrained and $\thC$ is essentially free, all the constraining power comes from the isotropic prior $p(\thobs) \sim \sin \thobs$, leaving both angles with large uncertainties.

Including the LIGO $\thobs$ posterior as a prior regulates the inference on $\thobs$ and pushes the inferred value lower, $(36 \pm 4)^\circ$ and $(27 \pm 6)^\circ$ for the fits without and with additional $L_X$, respectively.  Again the fit wit $L_X$ shows a larger uncertainty, as the additional component removes the Sedov constraint on $\thC$. 

The VLBI centroid provides a very strong constraint, inferring nearly identical values of $\thobs$ and $\thC$ regardless of whether the LIGO $\thobs$ prior is used. This is illustrative of the potential of combining GRB afterglow photometry, astrometry, and GWs for use in cosmology.
The centroid constraint shifts both $\thobs$ and $\thC$ downwards, keeping the $\thobs/\thC$ fixed to the light curve value, inferring $\thobs = (20.8 \pm 1.5)^\circ$ with only afterglow emission and $\thobs = (19.2 \pm 1.3)^\circ$ if there is an additional late time $L_X$.  The uncertainty is now marginally reduced in the presence of an additional $L_X$ instead of enhanced.  The VLBI centroid measurement prefers a viewing angle $\thobs \sim 20^\circ$, which the jet is free to choose with $L_X$ removing the pressure on $\thC$ to make up the late-time excess.  Without $L_X$, the late-time light curve and observed superluminal motion are in light tension, reflected in the slightly larger uncertainty.

This is reflected in the WAIC and LOO scores comparing the fits.  With no VLBI constraint, Table \ref{tab:Summary} shows a $<1 \sigma$ preference for additional $L_X$ with Gaussian jets.  If not for the centroid, a forward shock afterglow alone can comfortably fit the light curve data.  With the VLBI constraint included, Table \ref{tab:fullResults} shows a $\sim 2 \sigma$ (WAIC) and $\sim 1 \sigma$ (LOO) preference for additional $L_X$.

\subsection{What do Deep Newtonian and Poisson do?}

Table \ref{tab:Summary} includes the summarized fit results for a Gaussian jet with and without the Deep Newtonian corrections included (see Section \ref{sec:deepnewtonian}).   
The overall results are very similar, with all parameters agreeing within uncertainties. The jet-only fit produces a \emph{slightly} larger viewing angle, while the fit with an extra X-ray component reports a slightly larger $L_X$.
Both effects are to be expected, as the Deep Newtonian correction produces a slightly brighter Sedov phase afterglow which slightly eases the tension between the late time X-ray excess and the apparent superluminal proper motion.  Indeed, both fits without Deep Newtonian corrections prefer slightly larger values of $\thC$ and $\thobs$, although still well within the uncertainties of the corrected models.  In measuring the presence of an additional component in \gwgrb{}, not including the Deep Newtonian corrections would enhance the significance by only $0.1\sigma-0.3\sigma$ by the LOO and WAIC scores, respectively.

Table \ref{tab:Summary} also shows the fit results for a Gaussian jet using a Normal likelihood for all X-ray data.  The fit with additional $L_X$ finds near identical parameter values as the full model with Poissonian X-ray likelihoods.  In this case the low count observations, where the Poisson distribution is particularly important, are satisfied by the additional $L_X$ and the afterglow jet is left with very similar constraints as the in the full model.  

The fit with only an afterglow shows a smaller viewing angle and opening angle than its Poissonian counterpart, roughly halfway in between the values found in the full model with and without $L_X$.  When counts are low, the Normal distribution likelihood penalizes the model less for under-predicting the observations than the Poisson distribution.  This weakens the effect of the late time X-ray excess and allows the posterior to be pulled towards the preferred VLBI centroid solution, mildly reducing the tension.  In the measuring the presence of an additional component, using Gaussian statistics instead of Poisson on the X-ray observations lowers the significance by $0.1\sigma - 0.2\sigma$.

In this case, the underwhelming influence of both the Deep Newtonian regime and Poisson statistics is largely due to the overwhelming influence of the centroid observations, which drive the fit to configurations with $\thobs \sim 20^\circ$ that do not entirely account for all the late-time X-ray emission.  In general, we expect both additions to the pipeline will remain important to the accurate modelling of long-term GRB afterglows, particularly those without detected proper motion.

\section{Discussion}\label{sec:discussion}

\begin{figure}
    \includegraphics[width=\columnwidth]{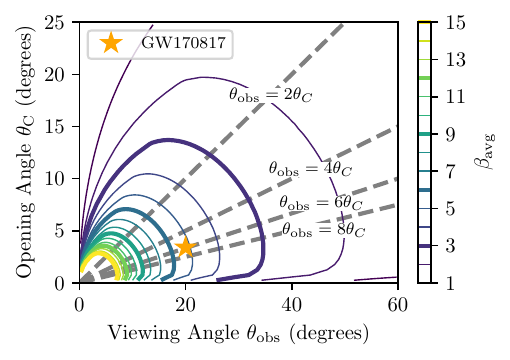}
    \caption{\label{fig:centroid_beta_map_gaussian}  The simple geometric constraints on a assuming a Gaussian jet from measurements of the early light curve slope and the centroid position.  The average apparent velocity of the afterglow centroid $\beta_{\mathrm{avg}}$ from launch to the jet break is shown (solid contours) as a function of $\thobs$ and $\thC$.  This quantity does not depend on the distance to the burst, the jet energy, nor the environment density. Lines of constant $\thobs / \thC$, which can be inferred from the early light curve slope, are shown by the grey dashed lines.  The orange star marks the values for \gwgrb{} under the Gaussian jet assumption, with a $\beta_{\mathrm{avg}} \approx 5$ and $\thobs / \thC \approx 6$.}
\end{figure}

The combination of photometric observations of a GRB's afterglow light curve and astrometric measurements of its proper motion can provide robust geometrical information on the jet width, as well as the orientation towards Earth.  Figure \ref{fig:centroid_beta_map_gaussian} demonstrates for Gaussian jets how these observations of a future GRB could constrain its geometry, and break the degeneracies present in each observation alone.  The pre-jet-break light curve evolution can provide a constraint on $\thobs/\thC$, placing the GRB along a straight line in $\thobs$--$\thC$ space.  Measuring $\beta_{\mathrm{avg}}$, the average apparent superluminal velocity to the jet break, provides an almost orthogonal constraint, identifying a curve of $\thobs$--$\thC$ space where the GRB must live.  The combination of these two constraints can pinpoint both $\thobs$ and $\thC$, subject to measurement uncertainties and assumptions of the jet model. 

The picture is complicated by more sophisticated jet models which admit more free parameters, and change the $\thobs/\thC$ constraint from the closure relations.  However, as shown in Figure \ref{fig:offset}, different jets seen at $\thobs \gtrsim 20^\circ$ display similar values of $\beta_{\mathrm{avg}}$, independent of the structures attempted.  The proper motion constraint is largely geometrical in nature and is likely more robust to jet structure assumptions than the early light curve slope. This is also reflected in Figure \ref{fig:centroid_beta_map_gaussian}, where for large viewing angles the contours approach vertical, indicating far stronger sensitivity to $\thobs$ than $\thC$.

\begin{figure}
    \includegraphics[width=\columnwidth]{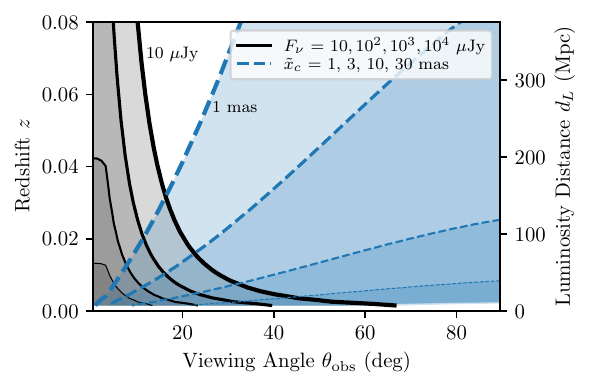}
    \caption{\label{fig:rangeFluxOffset}  The distance at which the radio flux and primary jet centroid offset at the jet break will take particular values as a function of viewing angle for a \gwgrb{}-like jet at $\nu$ = 5 GHz. Solid black contours show where the jet break flux flux $F_\nu = 10, 100, \dots$ $\mu$Jy.  Brighter jets are nearer and at lower $\thobs$, the shaded grey region shows where $F_\nu \ge 10$ $\mu$Jy, etc.  Dashed blue lines show where the centroid offset $\xcen = 1, 3, 10, 30$ mas at the time of the jet break.  Larger offsets occur for nearby jets with greater $\thobs$, the shaded blue regions shows where $\xcen \geq$ 1 mas, etc.  The overlapping shaded region shows roughly which \gwgrb{}-like jets will have detectable proper motion with VLBI.}
\end{figure}

Measuring proper motion from a GRB jet requires the motion be large enough to be observed  and the source be bright enough to be detected.  For jets observed at large inclination, the afterglow luminosity peaks at the time of the jet-break, which will be easiest time to detect proper motion. Figure \ref{fig:rangeFluxOffset} shows the distance at which a \gwgrb{}-like jet will display particular values of the radio flux and proper offset at the jet break.  At small viewing angles, the afterglow is bright and detectable to distances well beyond the LIGO volume, but the angular offset of the centroid is small.  At larger viewing angles the peak brightness drops precipitously, but the angular offset grows.  The ``sweet-spot'' where both effects may be detectable is a band of inclinations roughly centered on $\thobs \sim 16^\circ$.  At this inclination, a \gwgrb{}-like GRB at 100 Mpc will have a peak flux $\sim 30$ $\mu$Jy and show a proper centroid motion of $\sim$ 2 mas. 

GW events with EM counterparts are a promising standard siren cosmological probe.  Once a host redshift is known, the largest source of uncertainty in the GW measure of the luminosity distance is the unknown inclination angle \citep[e.g.][]{LIGOStandardSiren}.  With only the GW data and an EM-identified redshift, approximately 50 events are required to perform a 2\% measure of the Hubble constant \citep{Chen:2018aa}. The GRB afterglow may provide an independent, electromagnetic measure of the inclination of a binary neutron star merger event, breaking the distance-inclination degeneracy and reducing the number of events needed to perform precision GW cosmology.  Such a measure of $\thobs$ must be robust to the jet modelling systematics and likely will require detection and modelling of the afterglow proper motion, but would be a valuable tool for cosmology \citep{Gianfagna:2023ab}.

\citet{Hotokezaka:2019aa} fit the \gwgrb{} 300-day light curve with a semi-analytic afterglow model including the \citet{Mooley:2018} proper motion measurement and found $\thobs \approx 17^\circ$ 
for both Gaussian and power law jets.  \citet{Mooley:2022vs} performs a similar analysis with the 1000-day light curve, all radio VLBI measurements, and the \HST{} optical position and find $\thobs \approx 22^\circ \pm 3^\circ$.  Our results on the 2000 day light curve plus all centroid measurements find $\thobs$ around $21^\circ \pm 1.5^\circ$ for an afterglow jet alone and $19.3^\circ \pm 1.5^\circ$ for a jet plus an additional late-time X-ray component.  Our jet-only fits are generally consistent with the \citet{Mooley:2022vs} results, despite differences in the likelihood function (see Section \ref{sec:vlbidata}).  The larger uncertainties are likely due to their inclusion of the luminosity distance as a free parameter. The lower-inclination results from \citet{Hotokezaka:2019aa} used only a single apparent velocity measurement and did not include (as it had not yet been observed) the late X-rays which push the jet to larger opening (and hence, viewing) angles. Having only a single apparent velocity likely increases the risk of bias from the likelihood function, although the result is roughly consistent with our fits including a new X-ray component, which alleviates the late-time tension and allows for smaller inclinations.

\citet{Fernandez:2021to} tested several semi-analytic GRB jet models against the \gwgrb{} light curve and proper motion, finding $\thobs \sim 18^\circ$ to be consistent with observations.  \citet{Govreen-Segal:2023aa} fit an analytic model to the proper motion, informed by light curve fits, and found $\thobs = 19^\circ \pm 2^\circ$.  These are roughly consistent within uncertainties with both our results and \citet{Mooley:2022vs}, given the variety of methods and models employed. None of these works are able to fully reconcile the late time X-ray observations with the proper motion.

The late-time X-ray flux levels are in excess of what is expected from our jet models alone, under the condition they be consistent with the observed proper motion.  
\citet{Hajela:2022} reaches a similar conclusion, using both analytic and simulation-based jet models and a Poissonian treatment of the 900 and 1250 day X-ray observations. Figure \ref{fig:geometryInference} reflects this tension, in the $\thobs$ posterior distribution which favours anomalously high values in the afterglow-only fit.  If the jet model were doing a good job of predicting the data, one would expect perhaps a broad distribution in $\thobs$, which would be refined by the inclusion of the proper motion data.  Instead, we see a tight posterior at a large $\thobs$ which moves substantially when the proper motion data is included, with very little overlap with original distribution.  Quantifying this tension is the job of model comparison, in which we ultimately find that models with an additional X-ray source are only favoured at 1-2 $\sigma$.
Confirmation of whether a new component is appearing in \gwgrb{} will require more data at radio and X-ray wavelengths.

An unambiguous re-brightening of \gwgrb{} in radio or X-ray would likely be definitive evidence of a new component. If the slow fading continues, it will be difficult to differentiate between a new component, an unknown systematic in forward-shock afterglow models, or a statistical coincidence.  

Few GRB events have been followed up to the same degree as \gwgrb{}, but due to the spacetime volume observable with LIGO, future GW-GRBs will likely be nearby (compared to the EM-only GRB population) and have similarly substantial follow-up efforts.  For such events, observations of proper motion may be a crucial ingredient in determining the system inclination, and late-time analyses should be done with care, including the deep Newtonian regime and appropriate data likelihoods.

\section{Summary}\label{sec:summary}

GRB afterglows, being bright and beamed events, are imprinted with valuable information about the geometry and structure of the underlying relativistic jet and its progenitor system.  Decoding this information requires precise observations from across the electromagnetic spectrum and similarly precise modelling to be able to fit the data.  

We have incorporated several effects into a GRB afterglow model pipeline to improve its accuracy and constraining power.  These improvements are specifically targeted for the high-value GRB counterparts to GW events, which are likely to be seen at large inclination, be faint, and have substantial long term follow-up campaigns.  For faint X-ray observations, Poissonian statistics are important to reach accurate conclusions about the source flux.  At late times, the non-thermal synchrotron-radiating electrons become trans-relativistic and enter the \citet{SironiGiannios:2013} \emph{Deep Newtonian} regime: changing the evolution of the afterglow flux and peak frequency. Most relevant for geometrical constraints is incorporating measurements afterglow centroid proper motion.  The apparent superluminal velocity $\beta_{\mathrm{avg}}$ is highly sensitive to the viewing angle $\thobs$.  When $\thobs \gtrsim 20^\circ$, $\beta_{\mathrm{avg}}$ appears largely independent of the underlying jet structure.

We applied our model to the hallmark event \gwgrb{}, including the latest \Chandra{} X-ray observations, the radio VLBI measurements of the afterglow centroid position, and the recent HST constraint on the burst location.  The dataset is well decribed by a narrow jet of total kinetic energy $\sim 10^{52}$ erg, width $2^\circ-3^\circ$, viewed $\sim 20^\circ$ off the jet axis.  Models with an additional X-ray source emerging at late times provide a marginally better fit to the data, but are only favoured at 1-2 $\sigma$.  We do not find this convincing evidence for the presence of a new component in \gwgrb{}. 

Measurements of the afterglow centroid proper motion will be very valuable for future GRB and GW-GRB events.  Detectability, however, requires that an event has sufficiently small viewing angle so as to be bright enough for detection and that it has sufficiently large viewing angle that proper motion can be observed.  For events within $\lesssim 100$ Mpc both constraints can be met for a range of viewing angles centered on 20$^\circ$.  As these events will likely show a rising light curve, the easiest time to detect the centroid position is also the time with the greatest constraining power: as the light curve peaks at the jet break.

The new modelling components used in this paper, the deep Newtonian regime (optionally enabled as a new spectrum type) as well as displacement and size of the afterglow centroid, will be made public in \afterglowpy{} \texttt{v0.8.0}.

\begin{acknowledgments}
Research at Perimeter Institute is supported in part by the Government of Canada through the Department of Innovation, Science and Economic Development and by the Province of Ontario through the Ministry of Colleges and Universities. HJvE and LP acknowledge support by the European Union horizon 2020 programme under the AHEAD2020 project (grant agreement number 871158).
ET was supported by the European Research Council through the Consolidator grant BHianca (Grant agreement ID: 101002761). 
BO gratefully acknowledges support from the McWilliams Postdoctoral Fellowship at Carnegie Mellon University. 
\end{acknowledgments}

%

\vspace{5mm}
\facilities{HST, Swift(XRT), CXO}


\software{  \afterglowpy{} \citep{Ryan:2020aa},
            \texttt{astropy} \citep{astropy2013,astropy2018},
            \emcee{} \citep{Foreman-Mackey:2013aa},
            \texttt{matplotlib} \citep{Hunter:2007aa},
            \texttt{numpy} \citep{Harris:2020aa}
          }



\appendix

\section{Posterior Distributions For Full Model Fits}

\begin{figure}
    \centering
    \includegraphics[width=0.98\textwidth]{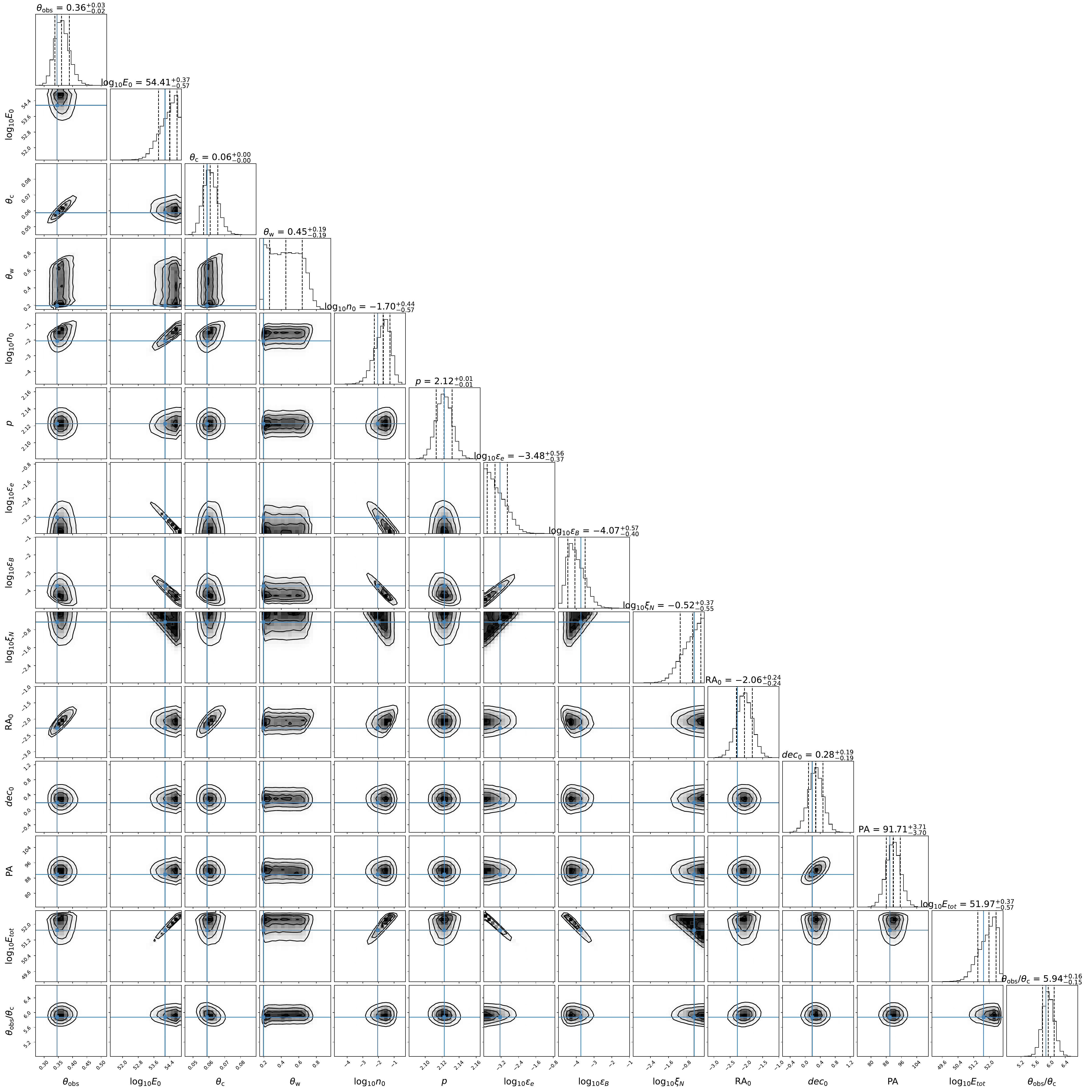}
    \caption{One- and two-dimensional marginalized posterior distributions for the ``full model'' fit to \gwgrb{} with only a Gaussian structured jet.  Written labels along the diagonal contain 68\% uncertainties, blue lines show the location of the maximum posterior probability density sample.
    \label{fig:MCMC_Gauss_NoLX_Corner}}
\end{figure}

\begin{figure}
    \centering
    \includegraphics[width=0.98\textwidth]{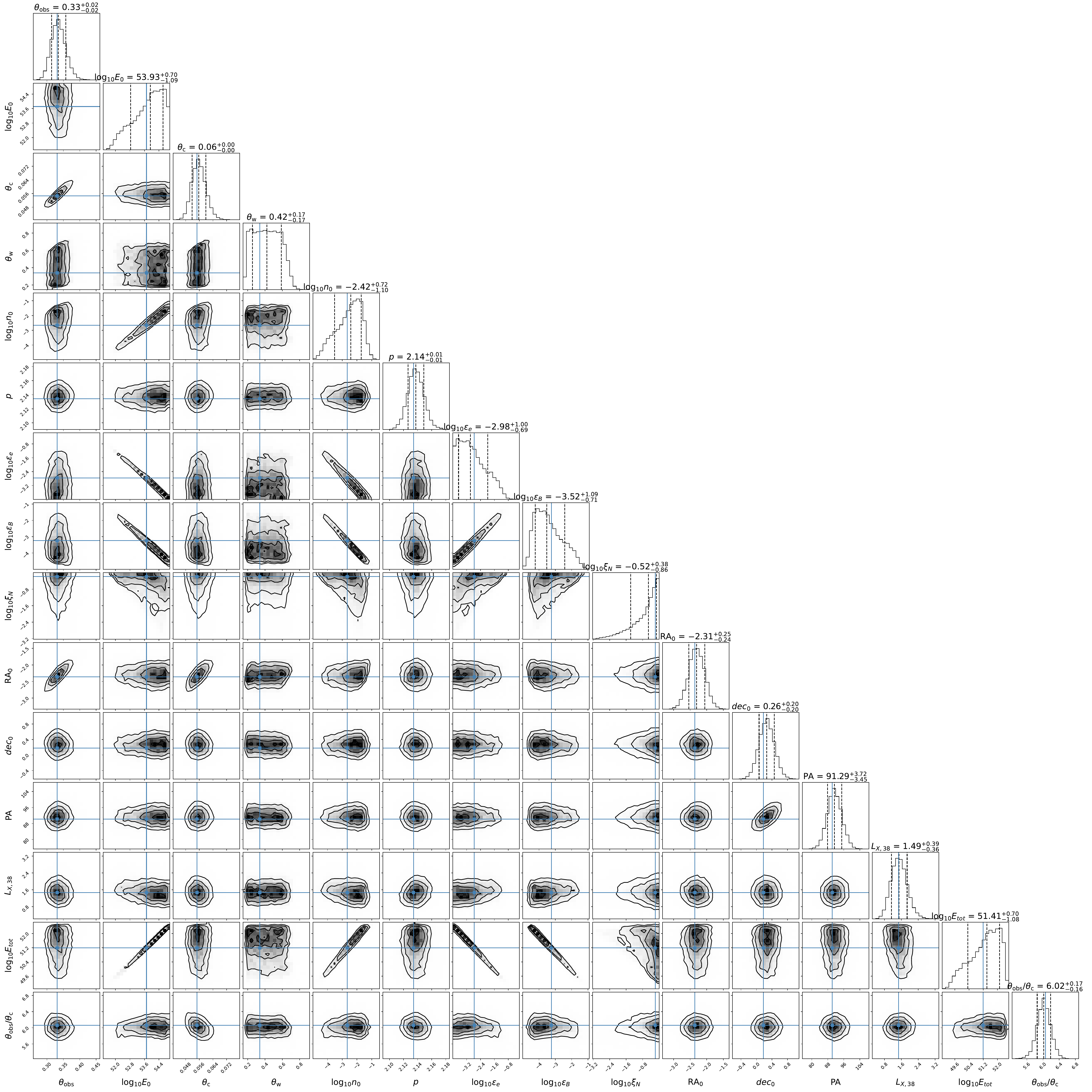}
    \caption{One- and two-dimensional marginalized posterior distributions for the ``full model'' fit to \gwgrb{} with a Gaussian structured jet and an additional constant luminous X-ray component $L_X$.  Written labels along the diagonal contain 68\% uncertainties, blue lines show the location of the maximum posterior probability density sample.
\label{fig:MCMC_Gauss_YesLX_Corner}}
\end{figure}

We show here the full posterior distributions for the Gaussian jet fits of the full model to \gwgrb{} as presented in Section \ref{sec:170817}.  Figure \ref{fig:MCMC_Gauss_NoLX_Corner} shows one- and two-dimensional representations of the posteriors of the fit with a Gaussian jet only.  Figure \ref{fig:MCMC_Gauss_YesLX_Corner} shows the same representation of the posterior for the fit with a Gaussian jet and an additional luminous X-ray component.


\bibliography{references}{}
\bibliographystyle{aasjournal}



\end{document}